\documentclass[11pt]{article}

\usepackage{amsfonts}
\usepackage{epsfig}
\usepackage{latexsym}
\usepackage{amsmath}
\usepackage{amssymb}
\usepackage{float}

\def\bequ{\begin{equation}}
\def\eequ{\end{equation}}
\def\barr{\begin{array}}
\def\earr{\end{array}}
\def\half{{1\over 2}}
\def\ben{\begin{equation}}
\def\een{\end{equation}}
\def\bena{\begin{eqnarray}}
\def\eena{\end{eqnarray}}

\def\spa#1{\phantom{\fbox{\rule[-#1cm]{0cm}{0cm}}}}

\def\b1{e^0}

\newcommand{\be}{\begin{equation}}
\newcommand{\ee}{\end{equation}}
\def\bea{\begin{eqnarray}}
\def\eea{\end{eqnarray}}

\def\half {{1 \over 2}}

\def\be{\begin{equation}}
\def\ee{\end{equation}}
\def\bea{\begin{eqnarray}}
\def\eea{\end{eqnarray}}

\def\lesssim{\mathrel{\hbox{\rlap{\hbox{\lower4pt\hbox{$\sim$}}}\hbox{$<$}}}}
\def\gtrsim{\mathrel{\hbox{\rlap{\hbox{\lower4pt\hbox{$\sim$}}}\hbox{$>$}}}}

\begin{document}
\title{{\LARGE \bf{Stationary Metrics and  Optical \\
      Zermelo-Randers-Finsler Geometry}}}
\author{
G. W. Gibbons$^{1,}$\footnote{G.W.Gibbons@damtp.cam.ac.uk}~, \ C. A. R. Herdeiro$^{2,}$\footnote{crherdei@fc.up.pt}~, \
      C. M. Warnick$^{1,}$\footnote{cmw50@cam.ac.uk} \ and M. C. Werner$^{1,3,}$\footnote{mcw36@cam.ac.uk}
\\
\\ {$^{1}${\em D.A.M.T.P.}}
\\ {\em  Cambridge University}
\\ {\em Wilberforce Road, Cambridge CB3 0WA, U.K.}
\\
\\ $^{2}${\em Departamento de F\'\i sica e Centro de F\'\i sica do Porto}
\\ {\em Faculdade de Ci\^encias da Universidade do Porto}
\\ {\em Rua do Campo Alegre, 687,  4169-007 Porto, Portugal}
\\
\\ $^{3}${\em Institute of Astronomy}
\\ {\em Cambridge University}
\\ {\em Madingley Road, Cambridge CB3 OHA, U.K.}}

\date{November 2008}       
 \maketitle

\begin{abstract}
We consider a triality between the Zermelo navigation problem, the
geodesic flow on a Finslerian geometry of Randers type, and spacetimes
in one dimension higher 
admitting a timelike conformal Killing vector field. From the latter viewpoint, the data of the Zermelo
problem are encoded in a (conformally) Painlev\'e-Gullstrand form of
the spacetime metric, whereas the data of the Randers problem are 
encoded in a stationary generalisation of the usual optical metric. We discuss how the spacetime viewpoint gives a simple and
physical perspective on various issues, including how Finsler geometries with constant flag curvature always map to conformally flat spacetimes and that the Finsler condition maps to either a causality condition or it breaks down at an ergo-surface in the spacetime picture. The gauge equivalence in this network of relations is considered as well as the connection to analogue models and the viewpoint of magnetic flows. We provide a variety of examples.
\end{abstract}

\newpage

\tableofcontents

\section{Introduction}
There has been a great deal of activity in theoretical physics
over the past few years  involving the properties of various spacetimes  
in four and higher dimensions. Apart from their obvious astrophysical
significance, there have been applications to string theory, gauge
theory (via the AdS/CFT correspondence), and a growing interest 
in analogue models. Most investigations have centred on particular
spacetimes, typically a solution of Einstein's equations
arising in supergravity theory. However, it is important
to distinguish particular properties from general ones. In other words,
we wish to see to what extent certain results obtained are due to the special
properties of the metric used and to what extent they are universal,
and valid for all metrics in a general class.  

In the case of {\it static metrics}  a convenient  tool for investigating
such questions is the well known  {\it optical metric} (see
\cite{Abramowicz} for an elementary account)  
which allows one to bring to bear on  
questions concerning  non-rotating  Killing horizons,  light bending etc.,  
the powerful  tools of Riemannian and Projective 
Geometry. For example,  the optical distance diverges as one 
approaches  a static Killing
horizon, and in this limit the metric is asymptotic to the metric on
hyperbolic space with the radius of curvature being given by the
inverse of the surface gravity of the horizon. 
One may use this fact to gain insight into 
``No Hair Theorems'' and the loss of information arising as 
matter falls into the associated black hole \cite{GibbonsWarnick}.
One may use the Gauss-Bonnet Theorem and information about the
curvature of the optical metric  to make general statements about
light bending \cite{GibbonsWarnick,Gibbons,GibbonsWerner}
and one may use the Projective Geometry to illuminate  
the dependence of light bending on the cosmological constant
 \cite{GibbonsWarnickWerner}.

This leads naturally to the question of what replaces 
the optical metric when one is dealing with {\it stationary metrics}
and can the associated geometric structures prove useful  
in dealing with rotating Killing horizons,  ergo-regions
and ergo-spheres, the existence  or otherwise of closed timelike
curves.  

The purpose of this paper  is to show how, in order to answer this question, one needs to pass from
Riemannian to  {\it Finsler}  Geometry, to obtain two spatial geometric objects,
a Randers and a Zermelo structure, related by  Legendre duality.
Taken with the original stationary spacetime,  one obtains a triple
of structures linked by a kind of  triality (see figure \ref{network}). This 
allows one to translate  a problem in one language to one of the other
two languages, often leading in this way to a dramatic
simplification. For example much effort has been expended on
understanding constant flag curvature for Randers-Finsler metrics
\cite{Shen}.
As we shall see, from the spacetime perspective the spacetime metrics
are rather simple: all constant flag curvature Randers-Finsler metrics map to conformally flat spacetimes. In particular, the principal example in \cite{Shen} 
lifts to flat  Minkowski spacetime in rotating coordinates, whereas the example in \cite{BaoShen} maps to an Einstein Static Universe in rotating coordinates.

The link between the three ideas is provided by a 1-form 
arising, in the spacetime picture,  from the cross  term $g_{0i}$  in the 
metric $g_{\mu \nu}$ of  the spacetime manifold $\{{ \cal M} , g \}$ and
associated with the {\it dragging of inertial frames}. When viewed
on the space  of orbits $\cal B$  of the time translation  Killing field  
 ${\partial / \partial t}$, this gives rise to a one-form $b_i$
which, with a metric $a_{ij}$,  endows ${\cal B}$ with a Randers structure,
a type of Finsler geometry which may be thought of physically in
terms of a fully non-linear and exact  version of 
{\it gravito-magnetism}.\footnote{In the case of ultra-stationary spacetime metrics this exact version of gravito-magnetism can be understood in terms of an equivalence of magnetic tidal tensors \cite{FilipeCosta:2006fz}.} Light rays in $\{ { \cal M} ,g\}$
project down on $\{{\cal B} ,a_{ij} ,b_i \}$  
to Finsler geodesics which in turn
 may be lifted to the cotangent bundle $T^\star{\cal B}$ to  
give a {\it magnetic flow}. Such flows have been extensively studied
by mathematicians  of late and, as we shall show,
some of their results have applications in the present setting. Moreover, we show that, in this magnetic picture, for all constant flag curvature Randers-Finsler 3-metrics, the magnetic field ${\bf B}$ is a Killing vector field of the Riemannian part of the geometry $a_{ij}$.

The alternative Zermelo viewpoint  works in terms of
a vector field or \textit{wind} ${\bf W}$, on ${\mathcal B}$, and a different metric
$h_{ij}$.  Light rays then project down on   $\{{\cal B} , h_{ij},
W^i \}$   to solutions of
the Zermelo problem: find the least time trajectory for a ship moving
with constant speed in a wind ${\bf W}$. Physically, this is where one
makes contact with fluid dynamical analogue models  
in which a fluid vortex, sometime thought of in terms of an  Aether,
gives rise to a Maelstrom-like  picture of a rotating black hole.
Currently, this idea, which can also work for a non-rotating black
hole, is  often associated with the use of Painlev\'e-Gullstrand coordinates. And here we show that, indeed, the Zermelo structure manifests itself in the spacetime picture by expressing the spacetime metric in a Painlev\'e-Gullstrand form.       

This paper is organized as follows. In section 2 we describe the network of relations: the triality between the Zermelo problem,  Randers-Finsler geometries and the null geodesic flow in conformally stationary spacetimes. We also discuss the magnetic flow viewpoint, the relation to analogue models and the gauge equivalence from  the spacetime picture. In particular, the condition of spacetime conformal flatness is explored in both the Randers and the Zermelo pictures, and the relation to Killing magnetic fields and constant flag curvature is established. In section 3 we provide a variety of examples of increasing complexity. We start with Langevin's rotating platform in Minkowski spacetime, which maps to a Randers-Finsler geometry of constant flag curvature, physically interpreted as a varying magnetic field in a negatively curved space, and a Zermelo problem of a constant angular velocity tornado (or Shen's rotating fish tank \cite{Shen}) in flat space. Let us note that, since the mapping between the spacetime and Randers picture can be seen as a temporal Kaluza-Klein reduction, this example is akin (in fact an analytic continuation) to the Melvin solution in Kaluza-Klein theory \cite{Dowker:1995gb}. Section \ref{magneticfields} explores, in contrast to the first one, several examples where the Randers picture is physically simpler than the Zermelo one, and in some cases, even simpler than the spacetime picture: constant or varying magnetic fields in $\mathbb{R}^3$ or $\mathbb{H}^2\times \mathbb{R}$. The particular example of section \ref{godelsection} allows us to relate the Finsler condition with isoperimetric inequalities, Novikov's action functional and ergodic theory. Section \ref{esu} gives another example of a Randers-Finsler geometry of constant flag curvature which is simply interpreted in the spacetime picture: a conformally flat spacetime -- the Einstein Static Universe (ESU) -- in rotating coordinates. We then consider a genuinely rotating ESU, 
 the Anti-Mach geometry, and its Randers and Zermelo pictures are exhibited. In section \ref{kerrsection}, we discuss the Randers and Zermelo pictures for the Kerr geometry. While these pictures are valid asymptotically, they break down at the ergo-sphere. To cover the ergo-region, we actually need to introduce co-rotating coordinates, and then the Randers/Zermelo pictures break down asymptotically. Using these co-rotating coordinates, we show that the Riemannian part of the Randers metric is hyperbolic near the horizon for non-extremal Kerr, in similarity with the general non-rotating Killing horizon. The Randers/Zermelo pictures of the near horizon extremal case, the NHEK geometry, are also considered and, in particular, the Zermelo picture turns out to be quite simple: an azimuthal wind in a deformed 3-sphere. We finish in section \ref{final} with some final remarks.

\section{A Network of Relations}
\subsection{The Zermelo problem as a Finslerian flow}
The Zermelo navigation problem \cite{Zermelo} is a time-optimal
control problem, which aims at finding the minimum time trajectories
in a Riemannian manifold $\{\mathcal{B},h\}$ ($\mathbb{R}^2$ with the
Euclidean metric in
\cite{Zermelo}) under the influence of a
drift (wind) represented by a vector field ${\bf W}$. Assuming a time
independent wind, Shen \cite{Shen} has shown that the trajectories which minimize
travel time are exactly the geodesics of a particular Finsler
geometry, known as \textit{Randers metric} \cite{Randers}, whose norm is given by 
\bequ
F(x,\mathbf{v})=\sqrt{a_{ij}(x)v^iv^j}+b_i(x)v^i \ , \ \ \ \  {\bf v}\in
T_x\mathcal{B} \ .  \label{randersle}\eequ
For the geodesics of
(\ref{randersle}) to solve the corresponding Zermelo problem, the
Randers data $\{a_{ij},b_i\}$ are determined in terms of the Zermelo data
$\{h_{ij},W^i\}$ as \cite{Robles}:
\bea
a_{ij} &=&{ \lambda  h_{ij} +W_i W_j \over \lambda ^2 }\ ,\qquad
\lambda =1- h_{ij} W^i W^j \ , \label{randers1}\\
b_i &=& -{W_i \over \lambda } \ , \qquad W_i = h_{ij} W^j \ .
\label{randers2} \eea
Note also that  
\ben
a^{ij} = \lambda \bigl (  h^{ij} - W^i W^j  \bigr )\ .
\een
It has also been shown \cite{Bao} that \textit{every} Randers metric $F$ arises as the solution to
 a Zermelo's problem of navigation, defined by the data
 $\{h_{ij},W^i\}$. The latter is defined by the former as: 
\bea
h_{ij} &=& \lambda  \bigl(  a_{ij} -b_i b_j \bigr ) \ ,\qquad \lambda 
=1- a^{ij} b_i b_j \ , \label{zermelo1}\\
W^i &=& -{b^i  \over \lambda } \ , \qquad b^i = a^{ij} b_j \ . \label{zermelo2}
\eea
Note also that 
\ben
h^{ij} = { \lambda a^{ij}  + b^i b^j \over \lambda ^2 } \ . 
\een
Thus, there is a natural identification of Randers metrics with
solutions to the Zermelo problem. It should be observed that 
\[ |W|^2\equiv h_{ij}W^iW^j=a_{ij}b^ib^j\equiv |b|^2 \ . \]
The \textit{Finsler condition} $|b|^2<1$ ensures that $F$ is positive and the metric is
convex, i.e. $\partial_{v^i}\partial_{v^j}\left({F^2}/{2}\right)$ 
is positive definite for all non-zero $\mathbf{v}$ \cite{Bao}.

\subsubsection{Zermelo/Randers duality via a Legendre transformation}
For a generic Finsler norm, the geodesics of $F$ are obtained by
taking $F$ as the Finsler Lagrangian. However, for a Hamiltonian
treatment, 
we need to consider a Lagrangian which is not homogeneous
of degree one in velocities, since the associated 
Hamiltonian would vanish. It is convenient to consider a Lagrangian of
degree two in velocities. Thus, since  $F(x,\mathbf{v}) $ is a  Finsler Lagrangian
which is homogeneous of degree  one, we define a Lagrangian by  
\ben
L= \half F^2\ .
\een
Hence, the Hamiltonian will  also be  of degree two in the  momenta
\ben
p_i = {\partial L \over \partial v^i}= F {\partial F \over \partial v^i}
\ .\een 
In fact, 
\ben
H(x,p)= L(x,\mathbf{v}) \ . 
\een
It is convenient to define a function $G(x,p)$ which
is homogeneous of degree one in momenta by
\ben
H= \half G^2\ .
\een 
Thus
\ben G(x,p) =F(x,\mathbf{v})\ .
\een
Specializing $F$ to be given by the Randers metric (\ref{randersle}), we have
\ben
p_i = F \bigl( n_i  + b_i \bigr)    \ ,\qquad n_i={ a_{ij}v^j \over 
\sqrt{a_{kr}v^kv^r} }  \ ; 
\een
since $a^{ij}n_i n_j=1$, 
we find that 
\ben
G= \sqrt{h^{ij}p_ip_j}  - W^ip_i \ , 
\een 
where $\{h_{ij}, W^i\}$  are the associated Zermelo data. Thus, the
Legendre transformation maps the Randers (Lagrangian) data to the
Zermelo (Hamiltonian) one.

Note that one may think of $G$ as the sum of two moment maps
\ben
G=G_0 + G_1 \ ,
\een
where
\ben
G_0=\sqrt{h^{ij} p_ip_j} \ , 
\een
generates the geodesic flow and
\ben
G_1=-W^i p_i \ ,
\een
generates the lift to the cotangent bundle of the
one parameter group of diffeomorphisms of the base manifold generated
by the vector field $-{\bf W}$. Observe that these two flows commute 
\ben
\{G_0,G_1\}=0\ ,
\een
if and only if ${\bf W}$ is a Killing vector field.
In the case that the base manifold is a sphere  and 
${\bf W}$   a Killing vector field, these are the flows
constructed by Katok \cite{Katok} and discussed  by Ziller \cite{Ziller}.
They correspond to rigid rotation on the sphere.

\subsection{Spacetime picture: Randers and Zermelo form of a stationary metric}
\label{stationary}
The geodesic flow of the Randers metric can be seen as the null
geodesic flow in a stationary spacetime with generic form
\ben
g_{\mu \nu} dx^\mu dx ^\nu  = -V^2 \bigl(dt + \omega _i dx^i \bigl )
^2  + \gamma _{ij} dx ^i dx ^j \ , \label{stationarymetric}
\een
so that
\ben
g_{ij}= \gamma _{ij} -V^2 \omega _i \omega _j\ .
\een
Fermat's principle arises from  the Randers structure given by
\bea
a_{ij}&=& V^{-2} \gamma _{ij}\ ,\label{staranders1}\\ 
a^{ij} &=& V^2 \gamma ^{ij}\ , \\
b_i& =& -\omega _i \ \label{staranders2}.
\eea   
Thus we call the form
\bequ
ds^2=V^2\left[-(dt-b_idx^i)^2+a_{ij}dx^idx^j\right] \ , \label{randersform} \eequ
the \textit{Randers form} of a stationary metric. Note that for $b_i=0$, $a_{ij}$ is the usual \textit{optical
  metric} (see \cite{Abramowicz} for an elementary introduction). 

Using (\ref{zermelo1}), (\ref{zermelo2}), the Randers data
are equivalent to a Zermelo structure of the form
\bea
h_{ij}&=& {1 \over 1 + V^2 g^{rs} \omega _r \omega _s } { g_{ij} \over
  V^2 } \ ,  \label{stazermelo1} \\
W^i&=& V^2 g^{ij}\omega _j \ , \label{stazermelo2}
\eea
where 
\bea
\gamma _{ij}&= & g_{ij} +V^2 \omega _i \omega _j\ \ , \\
\gamma ^{ij} &=& g^{ij} - { V^2 \omega ^i \omega ^j \over 1+ V^2
  g^{rs} \omega_r \omega _s } \ , 
\eea
and
\ben
\omega ^i = g^{ij} \omega _j \ ,\qquad W^i ={  V^2 \gamma ^{ij} \omega
  _j
\over 1-V^2 \gamma ^{rs} \omega _r \omega _s } \ ,
\een
\ben
1- V^2 \gamma ^{ij} \omega _i \omega _j= 
{1 \over 1+ V^2 g^{ij} \omega _i \omega _j}  \ .
\een
There is actually a more familiar way to extract the Zermelo data from
a stationary spacetime. Inverting (\ref{staranders1}),
(\ref{staranders2}) and using (\ref{randers1}), (\ref{randers2}) one
finds the spacetime metric (\ref{stationarymetric}) in the form
\bequ
ds^2=\frac{V^2}{1-h_{ij}W^iW^j}\left[-dt^2+h_{ij}(dx^i-W^idt)(dx^j-W^jdt)\right] \label{zermeloform}
  \ . \eequ
This is the \textit{Zermelo form} of a stationary spacetime. One
recognizes the metric in square brackets as having a 
Painlev\'e-Gullstrand form \cite{Painleve,Gullstrand}. Since in the spacetime picture of
the Zermelo problem/Randers flow one is interested only in the null
geodesic flow, the spacetime metric is only defined up to a conformal
factor. Thus, the Zermelo data encoded in a stationary spacetime can
be read off simply by writing it (conformally) in a Painlev\'e-Gullstrand form.

It should be noted, however, that for a given spacetime, an
apparently different Zermelo structure can be obtained by writing the
spacetime metric in Painlev\'e-Gullstrand \textit{coordinates}. It is helpful
to give an example. In these coordinates $(\tilde{t},r,\theta,\phi)$, the Schwarzschild metric is
\ben
ds ^2 = -d\tilde{t}^2 + (dr+vd\tilde{t})^2 + r^2 \bigl(d \theta ^2 + \sin ^2 \theta d
\phi ^2  \bigr)\ , \qquad 
v\equiv \sqrt{2M \over r} \ .
\een
One finds immediately that the Zermelo structure is
\ben
{\bf W}= -v {\partial \over \partial r } \ ,
\qquad h_{ij}dx^i dx^j= dr^2 + r^2 \bigl(d \theta ^2 + \sin ^2 \theta d
\phi ^2  \bigr)\ ,
\label{zersch1}\een
which is interpreted as a radial wind in flat space. On the other hand, the Zermelo structure derived by writing
the Schwarzschild solution in the usual Schwarzschild coordinates
$(t,r,\theta,\phi)$, in the form (\ref{zermeloform}), has no wind and the spatial metric is
the usual optical metric,
\ben  {\bf W}=0 \ , \qquad 
h_{ij} dx^i dx^j =  {dr ^2
  \over (1-v^2)^2 } + {r^2 \over 1-v^2}  \bigl(d \theta ^2 + \sin ^2
\theta d\phi^2\bigr) \ .
\label{zersch2}
\een
The two Zermelo structures (\ref{zersch1}) and (\ref{zersch2}) are
equivalent in the sense that the trajectories of one can be mapped to
the trajectories of the other one. This equivalence  is obvious in the
spacetime picture, since both structures correspond to the same
spacetime in different coordinates; but it is not obvious in the Zermelo
picture itself. In the Randers picture, on the other hand, the equivalence is again obvious, since $a_{ij}$ is, in both cases, the optical metric and $b_i$ differs only by a gauge transformation. This type of equivalence is one example of the large class of gauge
equivalences which will be present in our discussion of the
spacetime/Randers/Zermelo triangle and which we will explore more in
section \ref{sectiongauge}. In section \ref{kerrsection}, we will
briefly compare the analogue of these two Zermelo structures in the Kerr case.

Given the simple physical interpretation of the Zermelo problem, it
should not come as a surprise at this point that many analogue models
of black holes are naturally written in the Painlev\'e-Gullstrand form
(see, for instance
\cite{Volovik:1999fc,Unruh:2003ss,Visser:2007du}), as we shall now discuss briefly.

\subsubsection{Analogue Models: waves in moving media}
Metrics of Painlev\'e-Gullstrand form are frequently encountered 
in discussions of waves propagating in moving media and go back
at least to a paper of Gordon \cite{Gordon} on electromagnetic 
waves in a moving dielectric medium. Similar metrics have arisen when
discussing sound waves in a moving  compressible medium such as the atmosphere
\cite{White,MeyerSchroeter}. More recently attempts have been made to
construct analogue models of black holes \cite{Unruh,Barcelo:2005fc}
 and again similar metrics
are used.  Typically such metrics are taken to be accurate just to
quadratic order in  velocities and the underlying spatial metric to be
flat.
Thus one  considers a metric of the form
\ben
ds ^2= -c^2({\bf x}) dt^2 + \bigl ( d{\bf x} -{\bf W}  ({\bf x},t)  dt
  \bigr ) ^2\, , \label{analogue}
\een    
 where $c({\bf x} ,t) $ is the local light or sound  speed
and ${\bf W}({\bf x},t) $ the velocity  of the medium.
Both $c$ and ${\bf W} $ could, as indicated, be time dependent
but from now on we shall assume that both  are independent of
time. If $c$  and $ {\bf W}$ were constant, then (\ref{analogue}) would
be obtained
by performing a Galileo transformation on the flat metric
\ben
ds^2= -c^2 {dt^\prime } ^2 + ( d {\bf x} ^\prime )^2 \ , \qquad {\rm where}
\ \ \ 
t^\prime = t \ ,\qquad {\bf x} ^\prime  = {\bf x} - {\bf W} t\ . 
\een
The scalar wave equation for a scalar field $\Phi$, 
\ben
\left[{1 \over c^2} \left( {\partial \over \partial t^\prime} \right) ^2
- \left( {\partial \over \partial {\bf x} ^\prime} \right)^2
\right]\Phi=0 \ ,  
\een 
 becomes
\ben
\left[{1 \over c^2} \left( {\partial \over \partial t } + {\bf W}\cdot 
 {\partial \over \partial {\bf x} }    \right) ^2 
- \left( {\partial \over \partial {\bf x} } \right )^2\right] \Phi  =0
\ ,
\label{MorseIngard}  
\een
which is indeed the wave equation of (\ref{analogue}). One could adopt
(\ref{MorseIngard}) or some variant of (\ref{MorseIngard}) with lower order
terms even if $c$ and ${\bf W}$ depend on space and time. One would
 still find that the characteristics, i.e.  the \lq\lq rays \rq\rq \ 
 would be null geodesics of the metric (\ref{analogue}).

An interesting example is provided by a cylindrically  symmetric
vortex flow \cite{Leonhardt1,Leonhardt2}
for which, in suitable units, $c=1$ and  
\ben
{\bf W} = {\Omega \over \rho} \mathbf{e}_\phi  
\een
where $\rho= \sqrt{x^2 + y^2}$, $\tan \phi = y / x$  and 
${\bf e}_\phi$ is a unit vector in the $\phi$ direction.
The metric (\ref{analogue}) becomes
\ben
ds^2 = -dt ^2 + d\rho ^2 + \rho^2 \left( d \phi +{\Omega \over \rho ^2} dt
\right) ^2 + d z^2 \, . 
\een 
The metric is well defined and has Lorentzian signature for all 
$\rho>0$. The surfaces $\rho = {\rm constant}$ are timelike for all
$\rho >0$. Thus, there is no Killing  horizon, and hence no analogue
black hole,  in this metric
(in agreement with \cite{Visser2}). The surface 
$\rho = |\Omega | $ is a timelike ergo-sphere inside which the
Killing vector ${\partial/ \partial t }$ becomes spacelike. This
closely resembles the cylindrical analogues discussed by Zel'dovich when
he suggested the possibility of superradience for the Kerr black hole \cite{Zeldovich1,Zeldovich2}.

\subsection{The magnetic flow viewpoint}
\label{magnetic}
The Randers orbits are generated by the Finsler norm
(\ref{randersle}) taken as a Lagrangian:
$\mathcal{L}(x,\dot{x})=F(x,\mathbf{v})$. An immediate observation, already
pointed out in \cite{Randers} is that these can be interpreted as
motion in a magnetic field. Indeed, the Euler-Lagrange equations for
the Finsler norm are
\[ \frac{Du^i}{ds}=\tilde{F}^i_{\ j}u^i \ , \qquad \tilde{F}=db \ , \]
where $u^i=dx^i/ds$, and $s$ is arc length with respect to the
Riemannian metric $a_{ij}$. In terms of the standard magnetic potential
$A_i$ the orbits of a charged particle (mass $m$, charge $q$) can be
written, using the same parametrization, as
\[ \frac{Du^i}{ds}=\frac{q}{\sqrt{2m\epsilon}}F^i_{\ j}u^i \ , \qquad F=dA \ , \]
where the energy $\epsilon$ is
\[\epsilon=\frac{m}{2}a_{ij}\frac{dx^i}{d\tau}\frac{dx^j}{d\tau} \ ,
\]
and $\tau$ is proper time. Thus, the Randers 1-form $b$ relates to
the physical magnetic potential 1-form by
\bequ b=\frac{q}{\sqrt{2m\epsilon}}A \ . \label{randerstophysical}\eequ
Note, in particular, the dependence on the energy. We shall come back
to this point in the example of section \ref{godelsection}.

From a physicists' perspective, this is actually the most natural way to look at Randers metrics: as describing motion on curved manifolds with a magnetic field
\bequ
{\bf B}=B^i\frac{\partial}{\partial x^i} \ , \qquad B^i=a^{ij}(\star^{(3)} db)_j \ , \label{magfield}\eequ
denoting Hodge duality with respect to $a_{ij}$ by $\star ^{(3)}$.  From the perspective of section (\ref{stationary}), the spacetime
 geodesics, when projected down to the base manifold $\mathcal{B}$, with metric $a_{ij}$, are no longer geodesics; rather they   satisfy
 the equations of motion of a charged particle in a magnetic field,
 which is described by the Maxwell 2-form $ F$.  The particle paths may be lifted to the 
cotangent bundle of $T^\star \mathcal{B}$
 where they are referred to as \textit{magnetic flows}. They may be regarded as
a Hamiltonian system where the Hamiltonian coincides with
the usual  energy    
\ben
H= \half g^{ij}p_i p_j\ ,
\een
but where the symplectic form $\omega$  is modified
\ben
\omega = d p_i \wedge  d q^i + \frac{q}{2}F\ .
\een
This magnetic flow then uplifts to one dimension more (in a temporal version of the Kaluza-Klein picture) as the null geodesic flow discussed in section (\ref{stationary}).

Magnetic flows have been quite extensively studied
in the mathematics  literature. In particular, 
some of the phenomena to be discussed here, involving the breakdown of
the Randers structure and, at times, the transition  between the
behaviour with and without closed timelike curves in the spacetime
picture, has been noticed, albeit without that interpretation
\cite{BurnsPaternain}. To discuss them here it is useful to choose a
particular ``conformal gauge''.

\subsection{Spacetime versus magnetic flow pictures in the ultra-stationary gauge}
\label{ultrastationary}
With regard to the Randers/Zermelo structure, conformally related
spacetimes form an equivalence class. Thus, for this purpose, we can
always take the representative in this equivalence class to be
\textit{ ultra-stationary}, that is a spacetime $\mathcal{M}$ admitting 
an everywhere timelike Killing vector
field ${\bf K}$ with unit length, $g({\bf K}, {\bf K})=-1$. 
Such spacetimes  may be considered,  as a bundle over a base manifold $\mathcal{B}$
consisting of the space of orbits of the Killing vector field. These orbits,
sometimes called timelines,
may either be circles, $S^1$, in which case time is periodic, or lines
$\Bbb R$. In the former case there are obviously
closed timelike curves (ctc's). If the bundle is a trivial product, 
$\mathcal{M}\equiv S^1 \times \mathcal{B}$, these ctc's may be eliminated
by passing to a covering  space. 
Then the question is: do there exist ctc's which are homotopically trivial,
i.e. which may be shrunk to a point?
If the bundle is non-trivial, things may be more complicated. Spacetime 
may already be simply connected. 
In that case the ctc's may not be
eliminated by passing to a covering spacetime. This happens in
spacetimes of Taub-NUT type. If the bundle is non-trivial it will admit no 
global
section and hence no global {\it time coordinate} $t$, say, such that ${\bf K}t=1$. If the bundle is trivial then there will exist (many) global time coordinates $t$ but there may exist no global {\it time function}. A time function
may be defined to be a function which increases along every timelike curve.
Its level sets $t={\rm constant}$ must therefore be everywhere spacelike.  
The examples  that will be provided in section \ref{magneticfields}
are of this type: the bundle is trivial, a time coordinate exists, but
there is no global time function.
The line
element can be cast in the form (\ref{randersform}) with $V^2=1$
\bequ
ds^2=-(dt-b)^2+a_{ij}dx^idx^j \ , \label{usmetric} \eequ
where $b=b_idx^i$ and $x^i$ are coordinates on $\mathcal{B}$. The positive definite metric
$a_{ij}$ is the projection of the spacetime metric orthogonal to the
Killing vector field ${\bf K}= {\partial/ \partial t}$. The time coordinate $t$ is defined only up to a gauge transformation of the form $t \rightarrow t+ \psi(x)$ under which $
b \rightarrow b +d\psi$. The quantity $b_i$ may be regarded
as the pull-back to $\mathcal{B}$ of the Sagnac connection
\cite{Ashtekar:1975wt} which governs frame-dragging 
effects. It depends upon the choice of section of the bundle, however 
the pull-back of the 
curvature $\tilde{F}=db$ is gauge-invariant. 
 
Any timelike curve in $\mathcal{M}$ projects down to $\mathcal{B}$. We call its projection 
$\gamma$ and define its element of length with respect to the metric
$a_{ij}$ by $dl^2=a_{ij}dx^idx^j$.  The curves will be timelike as long as
\ben
dt> dl+b \label{in} \ .
\een
If the right hand side of (\ref{in}) is always positive
then $t$ is a time function. On the other hand, suppose that
$\mathcal{M}$ admits a closed timelike curve. Its projection on $\mathcal{B}$ will be a closed curve and moreover 
\ben
L(\gamma)+\int _\gamma b =0 \ ,
\een
where $L(\gamma)$ is the length of $\gamma$ with respect to the metric $a_{ij}$. Thus a \textit{sufficient} condition for the absence of ctc's is that
\ben
L(\gamma)>-\int _\gamma b \ , \label{cond}
\een
for all closed curves $\gamma$ in $\mathcal{B}$. Inequality (\ref{cond}) is also a \textit{necessary}
condition for the absence of ctc's, because if it is not true we can find a timelike curve for which the total
change in the coordinate $t$ vanishes, in which case it is a closed timelike
curve, or for which it is negative. 
In the latter case we can construct a closed timelike curve by
moving along an orbit of ${\bf K}$.   

The metric induced on the surfaces of constant $t$ is
\ben
ds^2|_{t=constant}=(a_{ij}-b_ib_j)dx^idx^j \ .
\label{tconstant}
\een
This depends on the choice of gauge. 
If for some choice of gauge it is positive definite, 
then the surfaces $t={\rm constant}$ will be spacelike and
and hence $t$ will be a time function. 
In that case no closed timelike curves are possible. But if
$(a^{ij}b_ib_j)|_{\mathcal{N}}=1$, where $\mathcal{N}$ is a
sub-manifold of the constant $t$ surfaces, (\ref{tconstant}) is not
positive definite on $\mathcal{N}$; this surface is either a
null or a singular surface where the Randers structure breaks down.

If $\mathcal{N}$ is regular, its existence implies the appearance of closed
null curves in spacetime. For this reason we call it a \textit{Velocity of
  Light Surface} (VLS), from the spacetime perspective. And if beyond $\mathcal{N}$ the metric induced on constant $t$ surfaces
becomes timelike, there will be closed timelike curves. Moreover, in
the absence of  horizons these can be extended all over the spacetime,
that is, they are \textit{naked} ctc's. In section \ref{magneticfields}
we will give four examples of this situation.

If $\mathcal{N}$ is singular, on the other hand, the ultrastationary
conformal gauge might not be the best gauge to provide a spacetime
interpretation. Multiplying the ultrastationary metric by a conformal
factor (that vanishes on $\mathcal{N}$) might provide a more physical
interpretation. This is illustrated by the examples in sections
\ref{langevin} and \ref{kerrsection}. In both cases, a spacetime
metric with an appropriate conformal factor renders the interpretation
of an \textit{ergo-surface} for $\mathcal{N}$, wherein
$\partial/\partial t$ becomes null. In this gauge $\mathcal{N}$ is
also a VLS, albeit of a different nature than that of the previous case.

\begin{figure}[h!]
\begin{picture}(0,0)(0,0)
\put(28,-14){Null Geodesic Flow in a}
\put(20,-29){Class of Stationary Metrics}
\put(65,-42){$V,\omega_i,\gamma_{ij}$}
\put(124,-120){Zermelo Problem}
\put(144,-135){$h_{ij},W^i$}
\put(-50,-113){Finslerian Flow in a}
\put(-38,-128){Randers Metric}
\put(-16,-140){$a_{ij},b_i$}
\put(-25,-70){Optical Form}
\put(71,-136){$_{(\ref{randers1}),(\ref{randers2})}$}
\put(71,-109){$_{(\ref{zermelo1}),(\ref{zermelo2})}$}
\put(-20,-80){$_{(\ref{staranders1})-(\ref{staranders2})}$}
\put(125,-70){Painlev\'e-Gullstrand Form}
\put(150,-80){$_{(\ref{stazermelo1})-(\ref{stazermelo2})}$}
\put(-105,-20){Temporal version of}
\put(-100,-35){Kaluza-Klein Form}
\put(-220,-120){Magnetic Flow $B^i$ in a}
\put(-221,-135){Riemannian Metric $a_{ij}$}
\put(-90,-133){$_{(\ref{magfield})}$}
\end{picture}
\centering\includegraphics[height=2in,width=6.5in]{{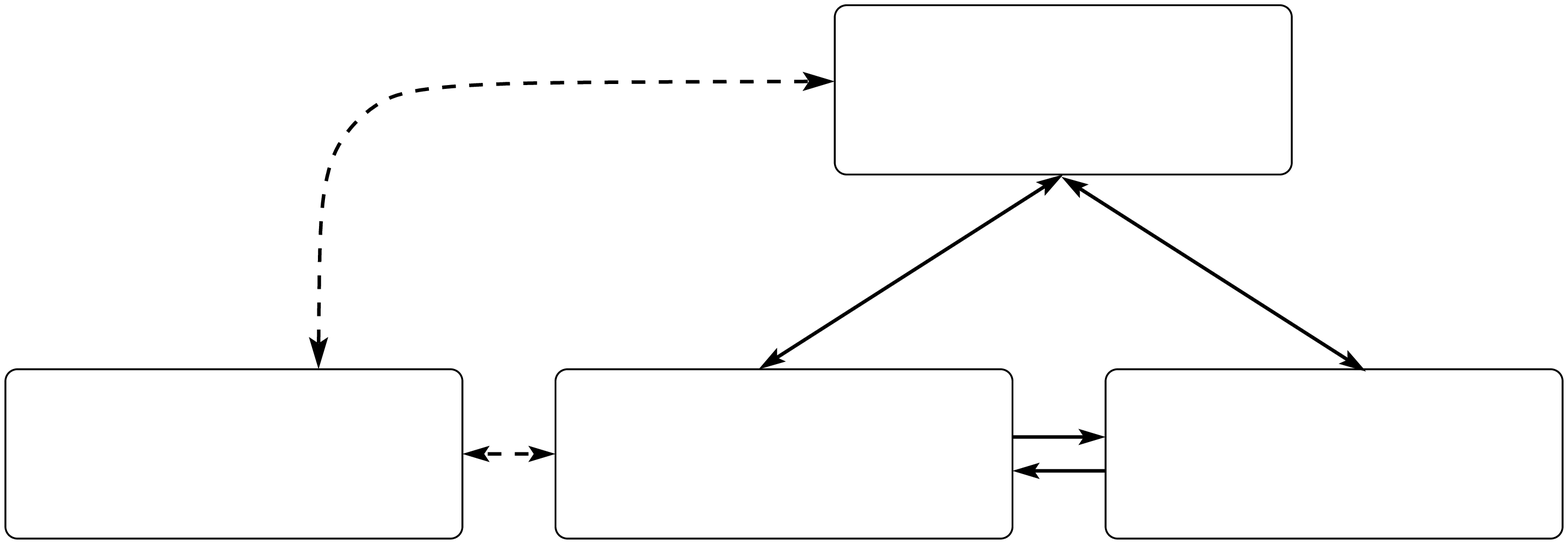}}
\caption{A network of relations. Alongside the arrows are the equation numbers which provide the respective map. The dashed lines represent interpretation relations rather than different frameworks.}
\label{network}
\end{figure}

\subsection{Gauge equivalence and Tensorial Relations}
\label{sectiongauge}
In figure \ref{network} we summarise the main relations discussed in the previous sections. In the network of relations described by figure \ref{network} it is possible that a complicated Randers picture is equivalent to a simple Zermelo picture or vice-verse. It is also possible that the spacetime description may illuminate the Randers or Zermelo pictures, or vice-verse. Or even that two Zermelo (or two Randers) pictures are equivalent, one being simple and the other involved. We shall now address the \textit{gauge equivalence} in the network. In the next section explicit examples will be provided.

The key observation here is that in the spacetime picture of the flows, we have a very large freedom to change the metric by diffeomorphisms (coordinate transformations) and conformal rescalings which do not affect the null geodesics. Because of this, we have access to a large class of transformations which map one problem into another, going far beyond the manifest symmetries of either of the lower dimensional viewpoints.

In the spacetime picture, we consider the null geodesics of a conformal class of metrics $[g]$. In order to pass to the Randers or Zermelo picture, we require that $[g]$ admits at least one timelike conformal Killing vector ${\bf K}$, satisfying:
\begin{equation}
\mathcal{L}_{\bf K} g = f g \ , \qquad g({\bf K},{\bf K})<0 \ ,  
\end{equation}
for some representative $g$ of $[g]$, where $f$ is any function on $\mathcal{M}$. The conditions are independent of the choice of representative $g$. Locally we may pick $g$ so that $f$ vanishes. We may choose coordinates so that ${\bf K} = \partial / \partial t$ and the conformal metric takes the form (\ref{usmetric}). This form of the metric is not uniquely determined: $b_i$ is determined only modulo an exact one-form so that the data
\begin{equation}
(b_i, a_{ij})\ ,\quad \mathrm{and}\quad (b_i + \partial_i \psi, a_{ij})\ ,
\end{equation}
arise from the same pair $([g], {\bf K})$. From the Randers point of view, this transformation of $b$ does not alter the geodesics, essentially adding an exact differential to the Lagrangian. In the spacetime picture this may be seen as a coordinate transformation to a new time coordinate $\tilde{t}$, as discussed in section \ref{ultrastationary}. Whilst this is a natural transformation from the point of view of the Randers picture, in the Zermelo case such a transformation gives rise to a complicated transformation involving both the wind {\bf W} and the spatial metric $h$.

In the case where the class $[g]$ admits more than one timelike conformal Killing vector, we are also free to make a different choice of ${\bf K}$. This may lead to very different structures on reduction to Randers or Zermelo data. As an example, we may consider the case where we have a Zermelo structure arising from a conformal Killing vector ${\bf K}$. If the Zermelo metric $h$ admits a Killing vector ${\bf \xi}$, then ${\bf K}+w{\bf \xi}$  may be seen to be a conformal Killing vector of $[g]$ and provided $w$ is sufficiently small this will remain timelike. After the reduction with this new conformal Killing vector, we see that the effect in the Zermelo picture is to introduce a `Killing wind'. Once again, this is a natural transformation in the Zermelo picture, but can be very complicated in the Randers picture, as shown in the examples of section \ref{langevin} and \ref{esu}.

 There are therefore essentially three possible transformations available to us, which relate Randers or Zermelo problems arising from the same geodesic problem in the spacetime picture.
\begin{itemize}
\item Change of coordinates in the spacetime picture independent of $t$. This descends to a change of coordinates on $\mathcal{B}$.
\item Shift of time coordinate in the spacetime picture: $t \to t+\psi$. This descends to a gauge transformation in the Randers picture $b_i \to b_i + \partial_i \psi$. In the Zermelo picture, this transformation is complicated.
\item Change of timelike conformal Killing vector. This includes the introduction of a \lq\lq Killing wind\rq \rq in the Zermelo picture, but may produce other transformations. In the Randers picture, this transformation is complicated.
\end{itemize}

We thus see that once we allow the more general class of transformations available in the spacetime picture, some problems which from the lower dimensional point of view seem inequivalent arise from choosing different coordinates on the same spacetime, or reducing along different directions within the same spacetime. Such transformations include, but are by no means limited to the manifest symmetries of the lower dimensional pictures, for example under change of coordinates on $\mathcal{B}$.

\subsubsection{Conformally flat spacetimes}
\label{weylflat}
Once we allow this larger set of transformations to act on Randers or Zermelo problems, we are faced with the question of how to identify two equivalent problems. The standard tensorial quantities of the lower dimensional pictures are not preserved under a general higher dimensional transformation, so we must look elsewhere for our answer. The solution is to consider the Weyl tensor of the higher dimensional metric, which is invariant under diffeomorphisms and conformal transformations. A case of particular interest is when the higher dimensional space is conformally flat, so that it may be brought by coordinate transformations to the form $C^2 \eta$, with $\eta$ the Minkowski metric. In this case the null geodesics are simple to find.

If we assume that a spacetime metric in the Randers (\ref{randersform}) or Zermelo (\ref{zermeloform}) form is conformally flat, this imposes conditions on the relevant lower dimensional data.  In terms of the Randers data, for a \textit{four dimensional spacetime}, these conditions may be reduced to, using the analysis in \cite{Grumiller:2006ww},
\bequ
C^{\mu}_{\ \alpha\beta\gamma}=0 \  \ \stackrel{(\ref{randersform})}{\Leftrightarrow} \ \left\{ \barr{c}
R_{ij}-\frac{1}{3}a_{ij}R=B_iB_j-\frac{1}{3}a_{ij}B^2 \ , \spa{0.3cm}\\
\nabla_{(i}B_{j)}=0 \ , \earr \right. \label{Weyltrivial}\eequ
where the tensorial quantities on the right hand side of the
equivalence are all computed from the metric $a_{ij}$, and ${\bf B}$
is defined by (\ref{magfield}), i.e the magnetic field in the magnetic
flow interpretation of the Randers picture. Thus, a necessary (but not
sufficient) condition for the spacetime Weyl triviality of a Randers
structure is that the magnetic field (\ref{magfield}) is a Killing
vector of the metric $a_{ij}$. This gives a simple test for examining
the Weyl triviality of an apparent complex Randers structure. Examples of
this will be given in sections \ref{langevin} and \ref{esu}. In section \ref{magneticfields} we will give examples of both Killing  and non-Killing magnetic fields corresponding, in the spacetime picture, to non-conformally flat geometry.

It is also worthwhile to calculate the Weyl tensor for a metric in the Zermelo or Painlev\'e-Gullstrand form. To do so it is convenient to first introduce an orthonormal basis $\{\bar{e}^a\}$ of forms for the Zermelo spatial metric $h$ together with the dual basis of vector fields $\{\bar{\mathbf{e}}_a\}$. We may then use the following orthonormal basis of forms
\begin{eqnarray}
{e}^0&=& dt  \ , \nonumber \\
{e}^a&=& \bar{e}^a + W^a dt \ , 
\end{eqnarray}
for the space-time metric in Zermelo form, where $W^a$ are the
components of the wind vector in the basis  $\{\bar{\mathbf{e}}_a\}$,
$a=1\dots n$. Letting $\nabla_a$ denote the Levi-Civita connection of $h$ in the given basis, we find that the vanishing of the Weyl tensor is equivalent to the following conditions on ${\bf W}$ and $h$
\begin{eqnarray}
0 &=& \frac{1}{n(n-1)} \left(n{}^{(n)}R_{ab} - \delta_{ab}{}^{(n)} R \right) - \theta_{a e}\theta^e{}_b + \frac{\theta \theta_{ab}}{n-1}  + \frac{n-2}{n-1} \theta^e{}_{(a}\psi_{b) e} + \frac{\delta_{ab}}{n} \left( \theta_{e f} \theta^{e f} - \frac{\theta^2}{n-1}\right)\ , \nonumber \\
0&=& \left(\delta_b{}^e \delta_a{}^f \delta_c{}^g -\delta_b{}^g \delta_a{}^f \delta_c{}^e -\frac{1}{n-1}[\delta_{ac}\delta_b{}^e \delta^{fg} -\delta_{ac}\delta_b{}^g \delta^{fe} +\delta_{ab}\delta_c{}^e \delta^{fg} -\delta_{ab}\delta_c{}^g \delta^{fe}]\right) \nabla_e \theta_{fg}\ ,  \nonumber \\
0&=& {}^{(n)}C_{abcd} + 2 \theta_{a[c}\theta_{d]b} - \frac{2 \theta}{n-2} (\delta_{a [c} \theta_{d] b} - \delta_{b[c} \theta_{d] a}) + \nonumber \\ && \frac{2}{n-2}(\delta_{a[c}\theta_{d]e}\theta^e{}_b - \delta_{b[c}\theta_{d]e}\theta^e{}_a) + \frac{2(\theta^2-\theta_{ef}\theta^{ef})}{(n-1)(n-2)} \delta_{a[c}\delta_{d]b}\ ,  \label{zermweyl}
\end{eqnarray}
where we have introduced the quantities
\begin{equation}
\psi_{ab} = \nabla_{[a}W_{b]}\ , \qquad \theta_{ab} = \nabla_{(a}W_{b)}\ , \qquad \theta = \theta^a{}_a \ , 
\end{equation}
and all indices are raised and lowered with $\delta$. The curvatures ${}^{(n)}C_{abcd}, {}^{(n)}R_{ab}, {}^{(n)}R$ refer to the metric $h$.

In the case where the ultra-stationary metric is in fact ultra-static, i.e.\ ${\bf W}=0$, we quickly deduce that the Weyl tensor vanishes if $h$ is both Einstein and conformally flat, which imply that $h$ must in fact be of constant sectional curvature. In other words the only conformally flat ultra-static spacetimes are Minkowski space, $\mathbb{R}\times S^n$ or $\mathbb{R}\times \mathbb{H}^n$, as shown in \cite{Brown:1982hb}.

\subsubsection{Constant Flag Curvature}

A class of spaces of particular interest in the study of Randers geometry are the spaces of `constant flag curvature' which generalise the concept of space of constant sectional curvature to the Finsler regime. It has been shown \cite{Bao} that the Randers metrics of constant flag curvature $K$ have corresponding Zermelo data satisfying
\begin{itemize}
\item $h$ has constant sectional curvature $K+\frac{1}{16}\sigma^2$;
\item ${\bf W}$ is a homothety of $h$, i.e.
\begin{equation}
\mathcal{L}_{\bf W} h = -\sigma h \ , 
\end{equation}
for a constant $\sigma$.
\end{itemize}
Substituting $\theta_{ab} = -\sigma \delta_{a b}$ into equations (\ref{zermweyl}) one quickly finds that all the terms not involving curvatures of $h$ cancel. Since a metric of constant curvature is necessarily both conformally flat and Einstein, the spacetime corresponding to a Randers metric of constant flag curvature is conformally flat.

We conjecture that the converse also holds, in other words we may characterise the Randers metrics of constant flag curvature as those giving rise to a conformally flat spacetime of the form (\ref{randersform}). Clearly if we allow ourselves the full freedom to make gauge transformations and change
the timelike conformal Killing vector $\mathbf{K}$ then any conformally flat spacetime
may be reduced to a Randers structure of constant flag curvature by taking $\mathbf{K} =
\partial/\partial t$ in standard Minkowski coordinates, which gives the trivial Randers
structure of $a_{ij}=\delta_{ij}, b_i=0$.

If we restrict ourselves to only allowing gauge transformations, so that we have no
freedom to change $\mathbf{K}$ then there are more possibilities. In order to classify
the possible Randers structures arising from a conformally flat spacetime, we must
classify the timelike conformal Killing vectors of Minkowski space. Any such vector
generates a conformal transformation and so may be written as a linear combination of
generators of the conformal group. These generators are:
\begin{eqnarray}
\mathbf{p}_\mu &=& \frac{\partial}{\partial X^\mu} \ , \nonumber \\
\mathbf{j}_{\mu \nu} &=& X_\mu\frac{\partial}{\partial
X^\nu}-X_\nu\frac{\partial}{\partial X^\mu}\nonumber \ , \\
\mathbf{d} &=& X^\mu \frac{\partial}{\partial X^\mu} \nonumber \ , \\
\mathbf{k}_\mu &=& 2 X_\mu X^\nu \frac{\partial}{\partial X^\nu} - X^2
\frac{\partial}{\partial X^\mu} \ .
\end{eqnarray}
We claim that any conformal Killing vector which is timelike in some region of Minkowski
space may be brought to the following form by a conformal transformation:
\begin{equation}
\mathbf{K} = (\mathbf{p}_t - \kappa \mathbf{k}_t) + \omega^{ij} \mathbf{j}_{ij} + v^i
(\mathbf{p}_i + \kappa \mathbf{k}_i) \ ,
\end{equation}
where $\kappa, \omega^{ij}, v^i$ are constants which we require to be such that the
vector remains timelike in some region. For each such $\mathbf{K}$, there is a choice of gauge where the Zermelo structure is of the following type:
\begin{itemize}
\item $h$ is a metric of constant sectional curvature $\kappa$;
\item $\mathbf{W}$ is a Killing vector of $h$, determined by $\omega^{ij}, v^j$;
\end{itemize}
For example, in the case where $\kappa=0$, $\omega^{ij}$ give rise to a wind along the
rotational Killing vectors and $v^i$ along the translational Killing vectors. The $\kappa
\neq 0$ case is less clear cut, but in both cases the $\omega^{ij}$ give rise to winds
along the orbits of some $SO(3)$ subgroup of the full isometry group.

It would appear that we have not recovered all of the constant flag curvature manifolds
of \cite{BaoRoblesShen}, since they find in addition the possibility that $h$ is flat
and $\mathbf{W}$ is a homothety. This case may be seen in the Randers form to be gauge
equivalent to the hyperbolic metric with no magnetic field, so is included in our
classification above. In fact, if we wish to classify constant flag curvature metrics
modulo gauge transformations, our list above contains redundancies as any Killing wind
which arises from a hypersurface orthogonal Killing vector may be removed by a gauge
transformation.

\section{Examples}
We shall now provide the Randers and Zermelo picture of a number of stationary spacetimes and discuss some physical phenomena using the different viewpoints.

\subsection{A rotating platform on $\mathbb{M}^{1,3}$: simple Zermelo; non-trivial Randers}
\label{langevin}
\begin{figure}[h!]
\centering\includegraphics[height=3in,width=4.5in]{{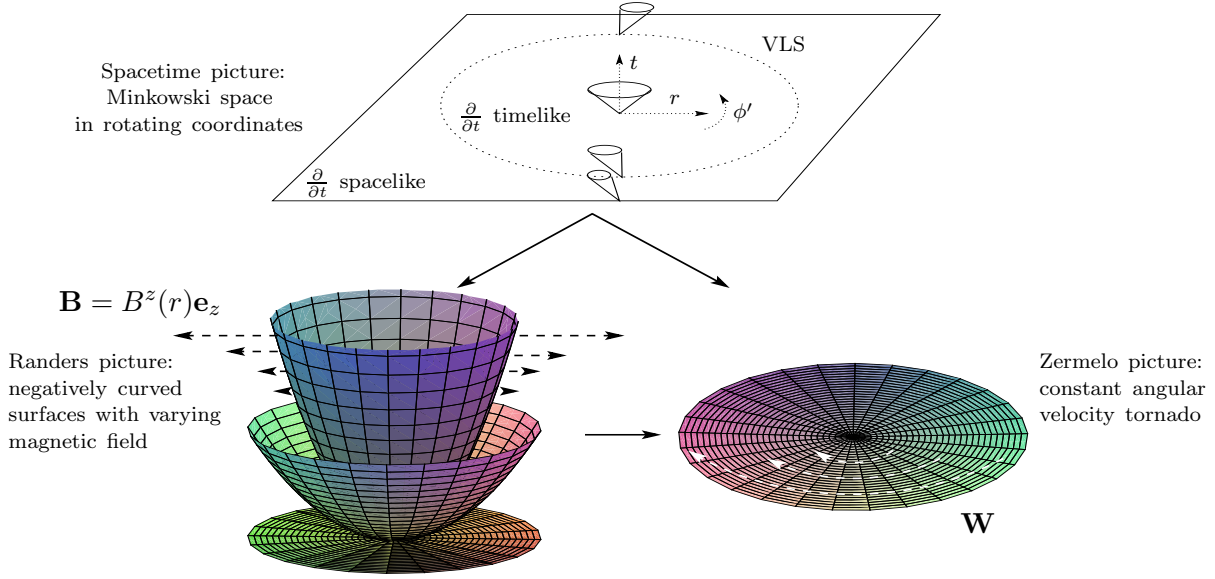}}
\begin{picture}(0,0)(0,0)
\put(-390,80){$_{{\rm Randers \ picture:}}$}
\put(-390,70){$_{{\rm negatively \ curved}}$}
\put(-390,60){$_{{\rm surfaces \ with \ varying}}$}
\put(-390,50){$_{{\rm magnetic \ field}}$}
\put(-371,100){${\bf B}=B^z(r){\bf e}_ z$}
\put(0,80){$_{{\rm Zermelo \ picture:}}$}
\put(0,70){$_{{\rm constant \ angular}}$}
\put(0,60){$_{{\rm velocity \ tornado}}$}
\put(-30,16){${\bf W}$}
\put(-355,190){$_{{\rm Spacetime \ picture:}}$}
\put(-353,180){$_{{\rm Minkowski \ space}}$}
\put(-365,170){$_{{\rm in \ rotating  \ coordinates}}$}
\put(-105,200){$_{{\rm VLS}}$}
\put(-155,193){$_{ t}$}
\put(-140,180){$_{r}$}
\put(-116,175){$_{\phi'}$}
\put(-220,175){$_{\frac{\partial}{\partial t} \ {\rm timelike}}$}
\put(-278,150){$_{\frac{\partial}{\partial t} \ {\rm spacelike}}$}
\end{picture}
\caption{Langevin's rotating platform in Minkowski space and the corresponding Randers and Zermelo pictures. We have chosen positive $\Omega$ to depict the tipping of the light-cones (spacetime picture) and the ``tornado'' direction (Zermelo picture). The direction of the magnetic field presented (Randers picture) is merely illustrative, since the embedding performed excludes the $z$ direction.}
\label{langevinnetwork}
\end{figure}

An elementary example is obtained by considering Minkowski space $\mathbb{M}^{1,3}$ in rotating coordinates. Let $\phi$ be the usual azimuthal cylindrical coordinate and $\phi'=\phi-\Omega t$ a co-rotating coordinate with a rigidly rotating platform (angular velocity $\Omega$). The spacetime metric, first considered by Langevin \cite{Langevin}, is then of the form (\ref{randersform}) with $V^2=(1-\Omega^2r^2)$ and  Randers data 
\bequ
b={\Omega r^2 d \phi' \over 1- \Omega ^2 r^2 } \ , \qquad 
a_ {i j}d x^i dx^j = {  dr ^2 + d z^2 
 \over 1-\Omega^2 r^2 } +
{r^2 d \phi'^2 \over \bigl ( 1-\Omega ^2 r^2 \bigr ) ^2  } \ . \label{randerslangevin}
\eequ
Let us consider briefly the $z=const.$ surfaces of this 3-geometry. For simplicity we take $\Omega=1$. Observe that, introducing $r=\sin\rho$, the geometry of these 2-surfaces has the simple line element 
\bequ ds^2=d\rho^2+A(\rho)^2d\phi'^2 \ , \label{2surface}\eequ
with 
\bequ A(\rho)=\tan\rho\sec \rho \ . \label{2surface2} \eequ
The surface (\ref{2surface}) has Gaussian curvature $K=-\ddot{A}/A$ (dot denotes $\rho$ derivative). Thus, our particular case has a negative, but not constant, Gaussian curvature, which diverges at $\rho=\pi/2$. Surfaces of the form (\ref{2surface}) may be embedded in a 3-dimensional Lorentzian/Euclidean space, with metric $ds^2=\pm dT^2+dX^2+dY^2$, using the embedding functions
\bequ X=A(\rho) \cos\phi' \ ,  \ Y=A(\rho) \sin\phi' \ , \ T=\int^\rho d\rho' \sqrt{\pm[1-\dot{A}(\rho')^2]} \ . \label{embed}\eequ
For the standard periodicity $\Delta \phi'=2\pi$, regularity at the origin requires $\dot{A}(\rho=0)=1$. Since, as we depart from $\rho=0$, $\dot{A}$ increases (decreases) for negatively (positively) curved surfaces, their embedding, if it exists, must be done in a Lorentzian (Euclidean) 3-dimensional space. Because of the $U(1)$ symmetry of (\ref{2surface}) one can display the embedding as the following surface
\[ T(\rho(R))=T(A^{-1}(R)) \ , \qquad  R\equiv \sqrt{X^2+Y^2}\ . \]
For instance, hyperbolic space, which has constant negative curvature, is described by $A(\rho)=\sinh\rho$. Thus, it is embedded in $\mathbb{M}^{1,2}$ as the surface
\[T(R)=\cosh(\sinh^{-1} R) \ . \]
For (\ref{2surface2}) it is straightforward to perform the integral (\ref{embed}); noting that 
\[A^{-1}(R)=\arcsin\left(\frac{\sqrt{1+4R^2}-1}{2R}\right) \ ; \]
  one can construct the embedding, which is displayed in figure
  \ref{langevinnetwork}. 

The impossibility of smoothly and isometrically embedding
  a 2-surface with a fixed point of a $U(1)$ symmetry, whose Gaussian curvature is everywhere negative, into
  Euclidean 3-space may be understood geometrically as follows. At every
  point the principal  curvatures are opposite in sign; thus any
  tangent plane cuts the surface into two parts, one lying on one side of the plane  and the other one lying on the opposite side.  If the smoothly embedded surface is such that it lies entirely on one side or the other of a
  complete plane (as in the case of a surface
  of revolution with a fixed point, such as the one displayed in figure
  \ref{langevinnetwork}) we can bring up this plane such that it touches this
  surface. At this point the plane coincides with the tangent plane,
  and hence we obtain a contradiction.

Rewriting the metric in a Painlev\'e-Gullstrand form one obtains the Zermelo data:
\bequ
{\bf W}=-\Omega \frac{\partial}{\partial \phi'} \ , \qquad h_ {i j}d x^i dx^j =   dr ^2 + d z^2 +r^2d\phi'^2 \ . \label{zermelolangevin}\eequ
This example illustrates the discussion of section \ref{sectiongauge} concerning ``Killing winds'': whereas the Zermelo picture (\ref{zermelolangevin}) retains the simplicity of the spacetime viewpoint -- a rigid rotation in flat space --, the Randers picture (\ref{randerslangevin}) appears to reveal a much more complex structure of a non-trivial one-form in a curved space. However, using the tensorial test (\ref{Weyltrivial}) one could also observe the Weyl triviality of the spacetime description of this Randers structure; in particular the magnetic field
\[{\bf B}= 2\Omega \frac{\partial}{\partial z} \ , \]
is indeed a (quite simple) Killing vector field of (\ref{randerslangevin}). Another instructive conclusion from this example is that the breakdown of the Randers/Zermelo picture for $r\ge 1/\Omega$  has a clear physical spacetime interpretation: the existence of a Velocity of Light Surface (VLS) at $r=1/\Omega$ and the consequent spacelike character of the $\phi'=const.$ surfaces for larger $r$. Outside this VLS timelike observers are obliged to move in the $\phi'$ direction.

In figure \ref{langevinnetwork} we summarize the network of relations for this example. In the three pictures we have shown the geometry of a $z=const.$ (and $t=const.$ in the spacetime picture) 2-surface. In the spacetime picture we have a VLS; both the Randers and the Zermelo picture cover only the region inside the VLS. In the Randers picture the 2-surfaces are negatively curved geometries; the outer surface depicted is the surface with constant negative curvature (and a flat surface) given for comparison; the norm of ${\bf B}$ is represented by the size of the dashed arrows; it diverges, together with the curvature, when the Randers picture breaks down. In the Zermelo picture the 2-surfaces are flat geometries; ${\bf W}$ describes a constant angular velocity ``tornado''. 

Let us finish this example with the comment that this Randers structure
is one of the examples discussed by Shen \cite{Shen} who envisages it
in terms of a fish moving inside a rotating fish tank. The Randers
structure has zero flag curvature and zero S-curvature. From the spacetime viewpoint this conclusion is somewhat trivial, since we are dealing with flat Minkowski spacetime.

\subsection{Magnetic fields on $\mathbb{R}^3$ and $\mathbb{H}^2\times
  \mathbb{R}$: simple Randers; non-trivial Zermelo}
\label{magneticfields}
We shall now consider four examples of magnetic fields: a constant and
a non-constant magnetic field in 3-dimensional flat space
$\mathbb{R}^3$ and a constant and a non-constant magnetic field in 2-dimensional
hyperbolic space times a flat direction $\mathbb{H}^2\times
  \mathbb{R}$. All of these examples are, in the spacetime picture,
  ultra-stationary metrics of the form (\ref{usmetric}), 
so that the discussion of section \ref{ultrastationary} applies.

\subsubsection{Heisenberg group manifold}
\label{heisenbergsection}
\begin{figure}[h!]
\centering\includegraphics[height=2.5in,width=4.5in]{{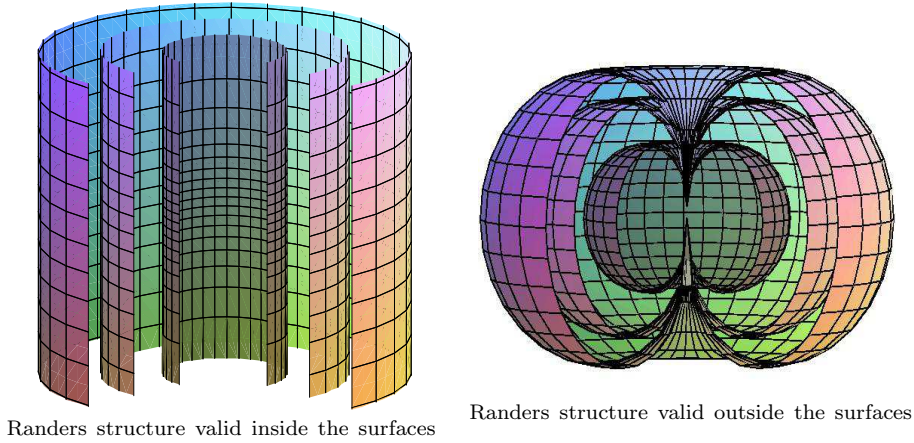}}
\begin{picture}(0,0)(0,0)
\put(-340,10){$_{\rm Randers \ structure \ valid \ inside \ the \ surfaces}$}
\put(-165,16){$_{\rm Randers \ structure \ valid \ outside \ the \ surfaces}$}
\end{picture}
\caption{Surfaces, in $\mathbb{R}^3$, where the Randers structure
  breaks down.  Left: Heisenberg example of section
  \ref{heisenbergsection} for $a=10/3,10/6,10/9$ (smaller to larger
  cylinder); the Randers structure is valid \textit{inside} the
  cylinder for each case. Right: St\"ormer example of section
  \ref{stormersection} for $\sqrt{\mu}=0.5,0.8,1$ (smaller to larger surface); the Randers
  structure is valid \textit{outside} the surface for each case.}
\label{randersheisto}
\end{figure}

A special case of the Som-Raychaudhuri spacetimes \cite{Som} is a
homogeneous manifold, which is (up to a trivial $z$-direction) the
group manifold of the Nil or Heisenberg group, the Lie group of the
Bianchi II Lie algebra. The spacetime metric is (\ref{usmetric}) with 
 the Randers structure
\bequ b=ar^2d\phi \ , \qquad a_{ij}dx^idx^j=dr^2+r^2d\phi^2+dz^2
\ . \label{constantmagnetic}\eequ
The corresponding magnetic field is ${\bf B}=a\partial / \partial z$; thus the
Randers picture has the simple physical interpretation of a constant
magnetic field in $\mathbb{R}^3$. This picture breaks down at
$r=1/|a|$ -- see figure \ref{randersheisto} -- which has the spacetime interpretation of being a VLS,
beyond which (i.e. for larger $r$), the integral curves of
$\partial/\partial \phi$ are closed timelike curves. Note also that this
Randers structure fails to obey the requirements (\ref{Weyltrivial}),
as expected, despite the fact that the magnetic field is actually a Killing vector field of (\ref{constantmagnetic}). 

The Zermelo structure, on the other hand, is more complex:
\bequ
{\bf W}=\frac{a}{1-a^2r^2}\frac{\partial}{\partial \phi} \ , \qquad h_ {i j}d x^i dx^j =(1-a^2r^2)[dr ^2 + d z^2 +(1-a^2r^2)r^2d\phi^2] \ . \label{zermeloheisenberg}\eequ

\subsubsection{Squashed $AdS_3$ and G\"odel spacetime}
\label{godelsection}
\begin{figure}[h!]
\centering\includegraphics[height=2.7in,width=5in]{{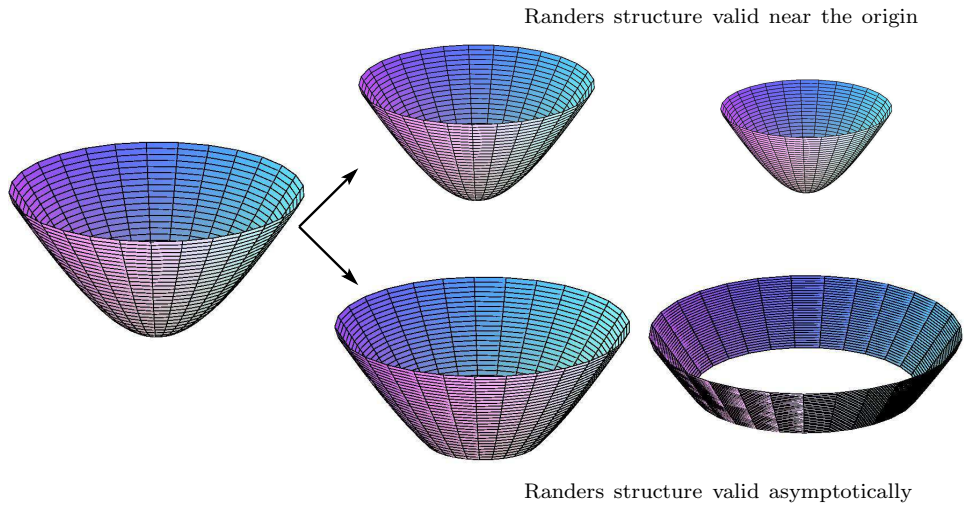}}
\begin{picture}(0,0)(0,0)
\put(-170,190){$_{\rm Randers \ structure \ valid \ near \ the \ origin}$}
\put(-170,10){$_{\rm Randers \ structure \ valid \ asymptotically}$}
\end{picture}
\caption{Region of $\mathbb{H}^2$ covered by the Randers
  structure in the examples of section \ref{godelsection} and
  \ref{cat}. The surface on the left is the embedding diagram of
  $\mathbb{H}^2$ discussed in section \ref{langevin}. As we increase 
  the constant magnetic field of section \ref{godelsection}, beyond
  the minimal value associated to $a=1$, the Randers structure
  breaks down outside a finite radius which decreases with increasing $a$
  (top sequence). As we turn on the asymptotically decaying magnetic
  field of section \ref{cat}, the Randers structure breaks down inside
  a finite radius which increases with increasing $\lambda$ (bottom sequence).}
\label{godelranders}
\end{figure}

$AdS_3$ is the group manifold of the Lie group $SL(2,\mathbb{R})$. Introduce the following two sets of 1-forms:
\bequ
\left\{ \barr{l} \sigma^0_L=d\phi+\cosh r dt \\ \sigma_L^1=\sin\phi
  dr-\sinh r\cos\phi dt \\ \sigma_L^2=\cos\phi dr+\sinh r\sin\phi dt
  \earr \ , \right. \qquad 
\left\{ \barr{l} \sigma^0_R=dt+\cosh r d\phi \\ \sigma_R^1=-\sin t
  dr+\sinh r\cos t d\phi \\ \sigma_R^2=\cos t dr+\sinh r\sin\phi d\phi
  \earr \ . \right. \eequ 
These are called \textit{left and right} 1-forms on
$SL(2,\mathbb{R})$, respectively. Both sets obey the Cartan-Maurer equations
\[d\sigma_L^\mu=\frac{1}{2}c^{\mu}_{\ \alpha\beta}\, \sigma_L^\alpha\wedge \sigma_L^\beta \ , \qquad
d\sigma_R^\mu=-\frac{1}{2}c^{\mu}_{\ \alpha\beta}\sigma_R^\alpha\wedge \sigma_R^\beta \ , \]
where $c^{\mu}_{\ \alpha\beta}$ are the structure constants of $sl(2,\mathbb{R})$, i.e
$c^{0}_{\ 12}=1=-c^{1}_{\ 20}=-c^{2}_{\ 01}$. The $AdS_3$ metric can be written in
terms of either set of forms; we write it as\footnote{For the $AdS$ space (i.e for $a^2=1$) to have unit ``radius'', a conformal factor of  $1/4$ would have to be included.} 
\bequ ds^2=-a^2(\sigma_R^0)^2+(\sigma_R^1)^2+(\sigma_R^2)^2 \
. \label{squashedads} \eequ
We have introduced a \textit{squashing} parameter $a$; $a^2=1$ corresponds
to the $AdS_3$ case. For $a^2>1$ we have a family of spacetimes with
ctc's of the G\"odel type \cite{Rooman:1998xf}. To identify the original
G\"odel universe \cite{Godel:1949ga} we introduce a
new time coordinate $t=2t'-\phi$ and the vector field
$U=\partial/\partial t'$. The Einstein tensor of (\ref{squashedads}) (with a
trivial flat direction) reads
\[
G_{\mu\nu}-a^2g_{\mu\nu}=4\frac{(a^2-1)}{a^2}U_{\mu}U_{\nu} \ , \]
which means it is a solution of the Einstein equations with a
cosmological constant $\Lambda=-a^2$ and a pressure-less perfect fluid
with density $\rho=4(a^2-1)/a^2$. G\"odel's original choice was to
take $\rho=|\Lambda|$ which corresponds to $a^2=2$, but all spacetimes
with $a^2>1$ are qualitatively similar. Introducing yet another time
coordinate $t=2at'$ the spacetime metric (adding a flat direction to
make it 4-dimensional) becomes of the form
(\ref{usmetric}) with the Randers structure
\bequ b=-2a\sinh^2\frac{r}{2}d\phi \ , \qquad
a_{ij}dx^idx^j=dr^2+\sinh^2r d\phi^2+dz^2 \ , \label{randersgodel}\eequ 
which can be interpreted as a constant magnetic field
${\bf B}=-a\partial/\partial z$ on $\mathbb{H}^2\times \mathbb{R}$. This
Randers structure breaks down at
\[ \tanh^2\frac{r}{2}=\frac{1}{a^2} \ , \]
being valid only for smaller $r$ -- see figure \ref{godelranders}. Thus, the Randers structure is valid
\textit{everywhere} if $a^2\le 1$. When it breaks down there is, in
the spacetime picture, a VLS,
such that the Killing vector field $\partial/\partial \phi$ becomes timelike
beyond it.\footnote{An illustrative diagram for the light cone
  structure of the G\"odel spacetime is given in \cite{HE} (and corrected in 
  \cite{Malament,OS0}). A similar light cone structure applies to the
  Heisenberg example of section \ref{heisenbergsection}. It is
  instructive to compare this light cone structure with the one illustrated
  in figure \ref{langevinnetwork}.}  

The Zermelo structure is, again, less straightforward to interpret:
\bequ
\barr{c}
\displaystyle{{\bf W}=\frac{a}{2[1+(1-a^2)\sinh^2\frac{r}{2}]}\frac{\partial}{\partial \phi}
 }\ , \spa{0.5cm}\\
    \displaystyle{h_{ij}dx^idx^j=\left(1-a^2\tanh^2\frac{r}{2}\right)\left(dr^2+\left(1-a^2\tanh^2\frac{r}{2}\right)\sinh^2r
    d\phi^2+dz^2\right)} \ . \earr \label{zermelogodel}\eequ

This squashed $SL(2,\mathbb{R})$ example allows us to make another 
application of the discussion of section \ref{ultrastationary}
together with \textit{isoperimetric inequalities}. For this purpose, 
reconsider equation (\ref{cond}), and suppose that the closed curve $\gamma$ spans a 2-surface
$\Sigma$. This will always be true if $\mathcal{B}$ is simply connected.
Using Stokes's theorem, condition (\ref{cond}) becomes 
\ben
L(\Sigma) > -\int _\Sigma \tilde{F} \ . \label{cond2}
\een
Using (\ref{randersgodel}) for a simply connected domain $\Sigma$ in the hyperbolic
plane $\mathbb{H}^2$, with area $A(\Sigma)$, condition (\ref{cond2}) becomes
\ben
L(\Sigma) >a A(\Sigma)\ .
\een
Now the isoperimetric inequality
of 
Schmidt \cite{Schmidt,Chavel} states that for any such domain (with Gaussian
curvature $K$)
\ben
L(\Sigma) > \sqrt{ 4 \pi A(\Sigma)  -KA^2(\Sigma)   } \ .\label{sch}
\een
In the Euclidean plane ${\Bbb E}^2$ the second term in the
square root would be absent; it arises from the negative curvature.
In fact for any two dimensional connected and simply connected domain with
 Gauss-curvature $K$ everywhere less or equal to $-\kappa$, Yau \cite{Chavel,Yau} has shown that
\ben
L(\Sigma) > \sqrt{\kappa}A(\Sigma). \label{yau}
\een
In general, the quantity
\ben
\inf L(\Sigma )/A(\Sigma),
\een
is called the \textit{Cheeger's constant} \cite{Cheeger}. 
Thus an inequality of the form (\ref{yau}) may be  called a generalised
isoperimetric or Cheeger type inequality.

Recall, from section \ref{ultrastationary}, that (\ref{cond2})
guarantees the absence of ctc's. Comparing with (\ref{yau}),  we see that if
$0<a \le 1$, (\ref{cond2}) holds for any $\gamma$ and hence there can be no
ctc's. Since the isoperimetric inequality is sharp, if $ a>1$ there
\textit{are} ctc's, in agreement with the spacetime picture.

The limiting case, case $a=1$, corresponds to the standard metric
on 
$\widetilde{ AdS_3} \equiv \widetilde{SL(2;{\Bbb R})} $,
the universal covering space of three-dimensional Anti-de-Sitter spacetime.
Of course it is not difficult to construct a time function directly
on $\widetilde{ AdS_3}$ but this  suggests an amusing  
reversal of the logic. Given that $\widetilde{AdS_3}$ admits no ctc's we {\sl deduce} that (\ref{yau}) holds 
for domains on $\mathbb{H}^2$. This leads to some apparently new
isoperimetric inequalities for 2-surfaces in  $\mathbb{H}^n_{\Bbb C}$, the complex hyperbolic ball
with its standard Bergmann metric.

The isoperimetric inequality and the existence or non-existence
of closed timelike curves in the spacetime picture is also closely related to
the behaviour of an action functional $S(\gamma)$
defined by Novikov \cite{Novikov} 
on $\Omega(x_1, x_2)$, the space of paths $\gamma$
joining two points $x_1$ and $x_2$ in $\mathcal{B}$.
The path pursued by a  particle moving with unit speed
in $\mathcal{B}$
under the influence of  a magnetic field $\tilde{F}$ 
may be found by extremizing this  action  functional which is
\ben
S(\gamma) = \int_\gamma (dl+b) \ , 
\een
among all curves $\gamma \in \Omega(x_1,x_2)$. Novikov proposes
using $S(\gamma)$ as a Morse function on $\Omega(x_1,x_2)$.
Unlike the length $L(\gamma)$ however, the functional $S(\gamma)$
need not be bounded below. This leads to difficulties with the Morse function
since the set in $\Omega(x_1,x_2)$ for which $S(\gamma) \le \epsilon$
is no longer relatively compact.

If this \lq \lq Arzela property\rq\rq  \cite{Novikov} does not hold, then
it is  not guaranteed that, for example,
there exists at least one critical path between two arbitrary points
$x_1$ and $x_2$ (see \cite{Novikov}). A simple example is provided
by a uniform magnetic field in the Euclidean plane. All classical paths are
circles with the Larmor radius (which is fixed for fixed physical
magnetic field and energy). If 
the distance between $x_1$ and $x_2$ exceeds twice the Larmor radius
there is no classical path between them. This is exactly what happens as we vary $a$ in the squashed
$SL(2,\mathbb{R})$ example:  $S(\gamma)$ will cease to be bounded
below as we cross $a=1$ towards $a>1$.

A last relation we would like to point out for this squashed
$SL(2,\mathbb{R})$ example is that the breakdown of the Randers
structure is associated to the \textit{transition from chaos to integrability}
for the magnetic flow. Take the Randers 1-form $b$ in terms of the
physical magnetic potential, $A$, for a particle of unit charge and unit
mass. Then, from (\ref{randerstophysical}),
$b=A/\sqrt{2\epsilon}$. Choosing the magnetic field $dA$ to be the
volume form on the unit radius (i.e Gaussian curvature $-1$)
hyperbolic plane $\mathbb{H}^2$, we recover the Randers structure
(\ref{randersgodel}) with 
\[ a=\frac{1}{\sqrt{2\epsilon}} \ . \]
It was shown in \cite{BurnsPaternain,Taimanov} that, under the conditions described,
the motion is completely integrable, in terms of real-analytic
integrals of motion, on the energy levels $2\epsilon<1$. For
$2\epsilon>1$, the magnetic flow is an Anosov flow (chaotic). Thus, the
transition from existence or otherwise of ctc's in the spacetime
picture and the breakdown or otherwise of the Randers structure in the
Randers picture, has yet another manifestation as a transition from a
non-ergodic to an ergodic motion in the magnetic flow picture.

\subsubsection{St\"ormer's problem}
\label{stormersection}
Consider the spacetime metric (\ref{usmetric}) with the Randers structure
\bequ b=-\mu\frac{\sin^2\theta}{r}d\phi \ , \qquad a_{ij}dx^idx^j=dr^2+r^2(d\theta^2+\sin^2\theta
  d\phi^2) \ , \label{stormerranders}\eequ
where $\mu$ is a constant. The corresponding magnetic field is a pure \textit{dipole} field:
\bequ {\bf B}=-\mu\left(\frac{2\cos\theta}{r^3}\frac{\partial}{\partial
  r}+\frac{\sin\theta}{r^4}\frac{\partial}{\partial \theta}\right) \
  . \label{dipolefield} \eequ
Notice that this is \textit{not} a Killing vector of $a_{ij}$ which immediately guarantees that the spacetime is not conformally flat and the Randers-Finsler geometry has not constant flag curvature.  From the form of ${\bf B}$ it follows that this Randers picture has the simple interpretation of a dipole
magnetic field on $\mathbb{R}^3$. Understanding the orbits of a
charged particle under the influence of such a magnetic field is often
called \textit{St\"ormer's problem} and it is of obvious interest for
understanding the Earth's magnetic field and the phenomena associated with it, such as the Polar Aurorae \cite{Stormer}.

The Randers structure (\ref{stormerranders}) breaks down at the surfaces 
\[r=\sqrt{|\mu|\sin\theta} \ , \]
which are represented in figure \ref{randersheisto}. In the spacetime
picture, these surfaces corresponds to a VLS beyond which
(i.e. for smaller $r$) the Killing vector field $\partial/\partial
\phi$ becomes timelike giving rise to ctc's. 

The Zermelo structure is
\bequ
\barr{c}
\displaystyle{
{\bf W}=\frac{\mu r}{r^4-\mu^2\sin^2\theta}\frac{\partial}{\partial \phi} \ ,} \spa{0.5cm}\\
\displaystyle{ h_ {i j}d x^i dx^j
=\left(1-\frac{\mu^2\sin^2\theta}{r^4}\right)\left[dr ^2 +
  r^2\left(d\theta^2
  +\left(1-\frac{\mu^2\sin^2\theta}{r^4}\right)\sin^2\theta
  d\phi^2\right)\right] \ .} \earr\label{zermelostormer} \eequ

It has been shown (see for instance \cite{Almeida}) that the St\"ormer problem is non-integrable, in a Liouville sense. Thus, one should not expect the geodesics of the spacetime metric (\ref{usmetric}) with (\ref{stormerranders}) to separate.

\subsubsection{G\"odel's cat}
\label{cat}
Cats always land on their feet. Even when dropped upside down and
without any external torque. Thus, by deforming their body, cats
undergo a rotation with zero angular momentum. 

Deformations generate rotation because, generically, they do not
commute.  The ratio between the rotation and the area of the
deformation can, when the latter tends to zero, be interpreted as
curvature. Thus, the optimisation problem of finding the most
effective rotation with zero angular momentum can be cast as motion in
a curved space. In \cite{avron} such problem was reduced to studying
the magnetic flow of the following Randers structure:
\bequ
b=-\lambda
a(r)\sinh r d\phi \ , \qquad  a_{ij}dx^idx^j=dr^2+\sinh^2r \,
{d\phi}^2+dz^2 \ , \qquad 
a(r)\equiv \frac{\cosh r-1}{\sinh 2r} \ ,  \eequ
where $\lambda$ is a constant. A trivial flat direction was added to the spatial metric to make it
three dimensional.  The interpretation of this Randers structure is
that of the non-constant magnetic
field
\[ {\bf B}=\frac{\lambda}{2\cosh^2 r}\frac{\partial}{\partial z} \ ,
\]
on a 3-geometry which is the direct product of the hyperbolic plane
with a flat direction $\mathbb{H}^2\times \mathbb{R}$. 

In the spacetime picture, the orbits are the null geodesics
of the 4-dimensional metric (\ref{usmetric}). Loosely speaking this is
an ``inside-out'' version of the G\"odel spacetime, albeit not
homogeneous. There is a VLS at \[ a(r)=\frac{1}{|\lambda|} \ , \]
wherein the Randers structure breaks down (it is only valid outside,
i.e for larger $r$ - figure \ref{godelranders}). Inside (outside) the VLS $\partial/\partial \phi$
is timelike (spacelike).

The Zermelo structure is:
\bequ
{\bf W}=\frac{\lambda a(r)}{\sinh r(1-\lambda^2
  a(r)^2)}\frac{\partial}{\partial \phi} \ , \qquad h_ {i j}d x^i dx^j =(1-\lambda^2
a(r)^2)[dr ^2 + d z^2 +(1-\lambda^ 2 a(r)^2)\sinh^2 rd\phi^2] \ . \label{zermelocat}\eequ

Let us close this section with an observation concerning the Zermelo structures (\ref{zermeloheisenberg}), (\ref{zermelogodel}), (\ref{zermelostormer}) and (\ref{zermelocat}). Generically, the 3-geometry $h_{ij}$ is conformal to the $t={\rm constant}$ sections of the spacetime picture geometry. The conformal factor that multiplies the latter to give the former is $(1-|b|^2)/V^2$, cf. (\ref{zermelo1}) and (\ref{randersform}). For ultra-stationary metrics $V=1$. Thus the surface at which the Randers structure breaks down becomes degenerate for the Zermelo geometry, even if it is non-degenerate in the spacetime picture. For the examples in this section, such surface is a smooth null surface in the spacetime picture, but becomes a singular surface (wherein the curvature blows up) for $h_{ij}$. For instance, the Ricci scalar of the 3-geometry (\ref{zermeloheisenberg}) is $\mathcal{R}=2a^2(7-5a^2r^2)/(1-a^2r^2)^3$.

\subsection{A rotating coordinate system in the ESU and the Anti-Mach metric}
\label{esu}
$S^3$ is the group manifold of the Lie group $SU(2)$. Introduce the following two sets of 1-forms:
\bequ
\left\{ \barr{l}  \bar{\sigma}_L^1=\sin\phi_2
  d\theta-\sin \theta \cos\phi_2 d\phi_1 \\ \bar{\sigma}_L^2=\cos\phi_2 d\theta+\sin \theta\sin\phi_2 d\phi_1 \\ \bar{\sigma}^3_L=d\phi_2+\cos\theta d\phi_1 
  \earr \ , \right. \qquad 
\left\{ \barr{l}  \bar{\sigma}_R^1=-\sin \phi_1
  d\theta+\sin \theta\cos \phi_1 d\phi_2 \\ \bar{\sigma}_R^2=\cos \phi_1 d\theta+\sin\theta\sin\phi_1 d\phi_2 \\ \bar{\sigma}^3_R=d\phi_1+\cos \theta d\phi_2
  \earr \ . \right. \label{formssu2} \eequ 
These are \textit{left and right} 1-forms on
$SU(2)$, respectively. Both sets obey the Cartan-Maurer equations
\[d\bar{\sigma}_L^i=\frac{1}{2}c^{i}_{\, jk}\bar{\sigma}_L^j\wedge \bar{\sigma}_L^k \ , \qquad
d\bar{\sigma}_R^i=-\frac{1}{2}c^{i}_{\, jk}\bar{\sigma}_R^j\wedge \bar{\sigma}_R^k \ , \]
where $c^{ijk}$ are the structure constants of $su(2)$, i.e
$c^{ijk}=-\epsilon^{ijk}$, where $\epsilon^{ijk}$ is the Levi-Civita tensor density and $\epsilon^{123}=1$. One can define the dual vector fields to $\bar{\sigma}_L^i$ and $\bar{\sigma}_R^i$, which we denote, respectively, by ${\bf k}_i^L$ and ${\bf k}_i^R$. To examine the symmetries of the geometries to be presented in this section, it is useful to notice
 that the left forms are \textit{right invariant} and the right forms are \textit{left invariant}:
\[ \mathcal{L}_{{\bf k}_i^R}\bar{\sigma}_L^j=0 \ , \qquad  \mathcal{L}_{{\bf k}_i^L}\bar{\sigma}_R^j=0 \ . \]
The Lie dragging of the left (right) forms along the left (right) vector fields is, on the other hand, given by
\[ \mathcal{L}_{{\bf k}_i^L}\bar{\sigma}_L^j=\epsilon^{ijk}\bar{\sigma}_L^k \ , \qquad  \mathcal{L}_{{\bf k}_i^R}\bar{\sigma}_R^j=-\epsilon^{ijk}\bar{\sigma}_R^k \ . \]

\begin{figure}[h!]
\centering\includegraphics[height=2.9in,width=3.9in]{{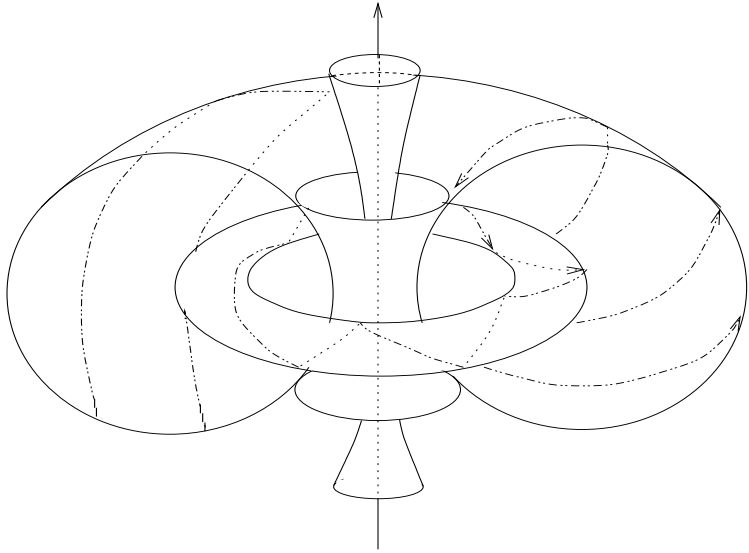}}
\begin{picture}(0,0)(0,0)
\end{picture}
\caption{Foliation of the 3-sphere by Clifford tori and the Killing wind corresponding, in the spacetime picture, to a rigidly rotating coordinate system. Adapted from \cite{Gibbons:1999uv} and inspired by the cover of \textit{Twistor Newsletter}.}
\label{esuwind}
\end{figure}

The Einstein Static Universe metric on $\mathbb{R}\times S^3$, with radius $R$,  can be written in terms of either set of forms; we write it as
\bequ ds^2=-dt^2+\frac{R^2}{4}\left[(\bar{\sigma}_R^1)^2+(\bar{\sigma}_R^2)^2+(\bar{\sigma}_R^3)^2\right] \
. \label{ESU} \eequ
Let us consider a rigidly rotating coordinate system in the ESU. Since this universe is finite, we expect that, for sufficiently small angular velocity, there will be no VLS and hence that the corresponding Randers/Zermelo structures will hold everywhere in $\mathcal{B}$. To see this explicitly consider the co-rotating coordinate $\phi'_1=\phi_1-\Omega t$, such that the rotation is along the integral lines of ${\bf k}_3^R$, i.e the Hopf fibres. The corresponding Randers structure is
\bequ
b=\frac{\Omega R^2}{4-(\Omega R)^2}\bar{\sigma}_R^3 \ , \qquad a_{ij}dx^idx^j=\frac{R^2}{4-(\Omega R)^2}\left[(\bar{\sigma}_R^1)^2+(\bar{\sigma}_R^2)^2+\frac{4(\bar{\sigma}_R^3)^2}{4-(\Omega R)^2}\right] \ , \label{randersrpesu}\eequ
where, the right-forms should be written in terms of $\phi_1'$ rather than $\phi_1$. This Randers structure breaks down if $R=2/|\Omega|$. Notice that this is now a condition on the metric parameters, rather than the equation of a surface. Thus if we take the angular velocity sufficiently small, $|\Omega|<2/R$, the Randers structure is valid all over $\mathcal{B}$. Physically this Randers structure is interpreted as a magnetic field along the Hopf fibre
\bequ{\bf B}=\frac{2\Omega}{R}\left[1-\left(\frac{\Omega R}{2}\right)^2\right] {\bf k}_3^R \ , \eequ
on a squashed 3-sphere (${\bf k}_3^R=\partial/\partial \phi_1'$). The Zermelo picture, on the other hand, retains the simplicity of the spacetime picture: a constant wind along the Hopf fibre of a \textit{round} $S^3$:
\bequ {\bf W}=-\Omega {\bf k}_3^R \ , \qquad h_{ij}dx^idx^j=\frac{R^2}{4}\left[(\bar{\sigma}_R^1)^2+(\bar{\sigma}_R^2)^2+(\bar{\sigma}_R^3)^2\right] \ . \label{zermelorpesu} \eequ
One can visualise this wind using the Clifford tori representation of the 3-sphere. Change from Euler $(\theta,\phi_1,\phi_2)$ to bipolar  $(\tilde{\theta},\tilde{\phi}_1, \tilde{\phi}_2)$ coordinates on $S^3$: $\theta=2\tilde{\theta}$,  $\phi_1=\tilde{\phi}_1+\tilde{\phi}_2$, $\phi_2=\tilde{\phi}_1-\tilde{\phi}_2$. The $\tilde{\theta}={\rm constant}$ surfaces of the 3-sphere become tori to which the wind is tangent since $\partial_{\phi_1}=(\partial_{\tilde{\phi}_1}+\partial_{\tilde{\phi}_2})/2$. This is represented in figure \ref{esuwind}.

Let us comment that the Finsler metric defined by (\ref{randersrpesu}) is equivalent to the family of Finsler metrics of constant, positive flag curvature on the Lie group $S^3$, constructed in \cite{BaoShen}. Explicitly, the parameter $K$ therein maps to 
\[ K=\frac{4}{4-(\Omega R)^2} \ . \]
So, the restriction considered in \cite{BaoShen}, $K>1$,  amounts to considering rotating coordinate systems in the ESU with $|\Omega|<2/R$, which is the condition, in the spacetime perspective, for $\partial /\partial t$ to be everywhere timelike and for the Randers structure to hold everywhere in $\mathcal{B}$. As in section \ref{langevin} and in agreement with the general discussion of section \ref{weylflat}, we again observe how the spacetime picture trivializes the construction of Randers-Finsler geometries with constant flag curvature.

There is actually an exact solution of Einstein's equations with a positive cosmological constant and incoherent matter describing a \textit{truly rotating ESU}, found by Osv\'ath and Sch\"ucking \cite{OS1,OS2}. This remarkable geometry, sometimes dubbed \textit{Anti-Mach metric}, lives on ${\Bbb R} \times S^3$
and  can be written\footnote{Our left invariant 1-forms $\bar{\sigma}_R^1$ and $\bar{\sigma}_R^2$ are exchanged as compared to the ones in \cite{OS2}. Since these can also be exchanged by making $k\rightarrow -k$ the solution presented here maps to the one in \cite{OS2} under this transformation.}
\ben
ds ^2 =-dt ^2 -R \sqrt{1-2k^2}\,  \bar{\sigma}_R^3 \, dt + { R^2 \over 4} \left[
 (1-k)(\bar{\sigma}_R^1 ) ^2 + (1+k) (\bar{\sigma}_R^2 ) ^2 + ( 1+ 2 k^2)
  (\bar{\sigma}_R^3)^2 \right]  \ ,
\label{antimach}\een
where $k$ is a constant, taken in the interval $0\le k^2 < { 1/ 4} $. The special case $k=0$ coincides
with the ESU, but described in
a frame of reference, i.e. a tetrad, which is  rigidly rotating
with respect to the standard static (and non-rotating)  frame. Actually this $k=0$ case is the rotating coordinate system just described with
\bequ
\Omega=-\frac{\sqrt{2}}{R} \ . \label{specialrp}\eequ
For direct comparison of the Randers/Zermelo structures that shall be obtained from the Anti-Mach metric with $k=0$ and the ones arising from the rotating platform with (\ref{specialrp}), it is convenient also to rescale the radius of the rotating platform, after implementing (\ref{specialrp}),
as $ R\rightarrow R/\sqrt{2}$. 
 
The geometry (\ref{antimach}) solves the Einstein equations with the energy-momentum tensor of 
pressure-free-matter,  with density 
$\rho$ and a cosmological constant $\Lambda$. 
The density, cosmological constant and radius $R$ are related by 
\ben
{4 \pi \rho \over \Lambda }= 1-4k^2 \ , \qquad \Lambda = {1\over R^2 (1-k^2)}\ . 
\een

In this geometry, the fluid is rotating with respect to the local inertial frame at some point $P$. Thus, 
it is a counter-example to (a version of) Mach's principle.\footnote{A similar property holds in G\"odel's spacetime \cite{Godel:1949ga}, but the latter has
ctc's and it is not finite. Thus this Anti-Mach metric is more physical and sharper (due to its finiteness) as a counter-example to Mach's principle
in general relativity.} Another interesting property of this geometry can be seen by noting that the 4-velocity of the matter is \cite{OS2}
\ben
{\bf u}= \sqrt{{ 1-2k^2 \over 2  (1-4k^2 )} } \, { \partial \over \partial t}
+ {1 \over R }\sqrt{{  2 \over  1-4k^2 } } \, {\bf k}_3^R\ . \een
Here and below ${\bf k}_3^R=\partial/\partial \phi_1$. The general metric is homogeneous, being invariant
under time translations, generated by $\partial / \partial t$,
and left translation of the 3-sphere  $S^3 $ generated by ${\bf k}_i^L$.
If $k\ne 0$, the fluid flow vector is {\sl not} a Killing vector field
and indeed, in   co-moving coordinates adapted to the flow, the metric
is not stationary, but rather it {\sl oscillates}  in co-moving
time.

The Randers structure for the Anti-Mach metric is
\bequ
b=-\frac{R}{2}\sqrt{1-2k^2}\, \bar{\sigma}_R^3 \ , \qquad a_{ij}dx^idx^j=\frac{R^2}{4}\left[(1-k)(\bar{\sigma}_R^1)^2+(1+k)(\bar{\sigma}_R^2)^2+2(\bar{\sigma}_R^3)^2\right] \ . \label{randersam}\eequ
This Randers structure never breaks down in the allowed range for $k$. Physically it is interpreted as a magnetic field along the Hopf fibre
\bequ{\bf B}=-\frac{2}{R^2}\sqrt{\frac{2(1-2k^2)}{1-k^2}} \, {\bf k}_3^R \ , \eequ
on a \textit{tri-axial} squashed 3-sphere. Observe that the magnetic field is \textit{not} a Killing vector field, immediately denouncing, by virtue of the discussion in section \ref{weylflat}, that the spacetime picture is \textit{not} a conformally flat geometry. The Zermelo picture, on the other hand, reveals a non-Killing wind, still along the Hopf fibre of a tri-axial squashed $S^3$:
\bequ{\bf W}=\frac{2\sqrt{1-2k^2}}{R(1+2k^2)} {\bf k}_3^R \ , \  h_{ij}dx^idx^j=\frac{(1+2k^2)R^2}{8}\left[(1-k)(\bar{\sigma}_R^1)^2+(1+k)(\bar{\sigma}_R^2)^2+(1+2k^2)(\bar{\sigma}_R^3)^2\right] \ . \label{zermeloam}\eequ
It is a simple matter to verify that (\ref{randersam})-(\ref{zermeloam}), with $k=0$, reduce to (\ref{randersrpesu})-(\ref{zermelorpesu}) with (\ref{specialrp}) and the appropriate rescaling of $R$.

\subsection{Kerr black hole and charged Kerr}
\label{kerrsection}
\subsubsection{Asymptotic Randers/Zermelo structure}
\label{kerrasymptotic}
We now consider rotating black holes. In Boyer-Lindquist coordinates $x^\mu=(t,r,\theta,\phi)$, the Kerr solution is given by the line element
\begin{equation}
ds^2=-\frac{\Delta}{\rho^2}\left(dt-a\sin^2 \theta d\phi\right)^2+\frac{\sin^2\theta}{\rho^2}\left((r^2+a^2)d\phi-a dt\right)^2+\frac{\rho^2}{\Delta}dr^2+\rho^2d\theta^2\ ,
\label{kerr1}
\end{equation}
where $m$ is the ADM mass and $a=j/m$, with $j$ the ADM angular momentum; moreover 
\[\Delta\equiv r^2-2mr+a^2 \ , \qquad \rho^2\equiv r^2+a^2\cos^2\theta \ . \]
A simple manipulation recasts the metric in the form
(\ref{randersform}) with $V^2=H/\rho^2$ and the Randers structure
\bequ
b=-\frac{2mra\sin^2\theta}{H}d\phi \ , \qquad  a_{ij}dx^idx^j=\frac{\rho^4}{H}\left(\frac{dr^2}{\Delta}+d\theta^2+\frac{\Delta\sin^2\theta}{H}d\phi^2\right) \ , \label{randerskerr2}
\eequ
where
\bequ H\equiv \Delta - a^2\sin^2\theta \ . \label{H} \eequ
The magnetic field that one derives from this Randers structure is
\bequ  {\bf B}=-\frac{2ma\sin\theta(r^2-a^2\cos^2\theta)}{\rho^6}\frac{\partial}{\partial \theta}-\frac{4mra\cos\theta\Delta}{\rho^6}\frac{\partial}{\partial r}\label{kerrmagnetic} \ . \eequ
One notes that, on the equatorial plane ($\theta=\pi/2$), the magnetic
field is purely in the $\theta$ direction; thus a particle with a
3-velocity on the equatorial plane will feel a Lorentz force on the
equatorial plane, as expected from the fact that this sub-manifold is
totally geodesic. Likewise, along the axis direction
($\theta=0,\pi$), the magnetic field is purely radial. Thus a particle
moving along the axis will feel no Lorentz force and therefore will
remain moving along the axis, as expected from the fact that this
sub-manifold is also totally geodesic in the full Kerr
metric. Observe also that, for large $r$, this magnetic field
approaches the pure dipole field (\ref{dipolefield}) with $\mu=2j$. 
The Finsler condition is given by
\[
|b|^2=\frac{(2mra\sin\theta)^2}{\Delta\rho^4}<1\ .
\]
This Randers structure is therefore only defined outside the ergo-region. To see this, note that 
\[  \rho^4\Delta-(2mra\sin\theta)^2=H((r^2+a^2)^2-a^2\sin^2\theta \Delta)\equiv C(r,\theta) \ , \] and at the ergo-sphere $H=0$. Hence the Finsler condition fails to hold exactly on the ergo-sphere. The spacetime interpretation of the breakdown of the Randers structure is very similar to the case of the rigidly rotating platform of section \ref{langevin} (but an inside out version): the ergo-sphere is a VLS in the sense that $\phi=const$ becomes a spacelike surface inside the ergo-region.

The Zermelo structure for the Kerr metric is, on the other hand,
\bequ \barr{c}
\displaystyle{{\bf W}=\frac{2marH}{C(r,\theta)}\frac{\partial}{\partial \phi} \ , \qquad  h_{ij}dx^idx^j=\frac{C(r,\theta)}{\Delta H}\left[\frac{dr^2}{\Delta}+d\theta^2+\frac{C(r,\theta)}{H\rho^4}\sin^2\theta d\phi^2\right]} \ . \earr \label{zermelokerr}\eequ
Observe that this Zermelo structure is qualitatively different from the `flow of space picture' obtained in \cite{Natario:2008ej}, by considering Painlev\'e-Gullstrand \textit{coordinates} for the Kerr metric based on zero angular momentum observers dropped from infinity; namely the wind herein is purely azimuthal whereas the `flow of space' in \cite{Natario:2008ej} has a radial component (as the wind in (\ref{zersch1})), a distinction which is compensated by the different space curvature.

Let us note that this analysis of Zermelo/Randers structure of the Kerr black hole generalizes quite easily to charged rotating black holes. The metric of the Kerr-Newman black hole of Einstein-Maxwell theory is obtained by simply
replacing
\[ \Delta \rightarrow \Delta + q^2 \ , \]
in (\ref{zermelokerr})/(\ref{randerskerr2}), where $q$ denotes the black hole charge. The
metric of the Kerr-Sen \cite{Sen:1992ua} black hole of low energy heterotic string theory is obtained, as observed in
\cite{Galtsov:1994pd}, simply by replacing
\[ \Delta \rightarrow \Delta + \frac{q m}{q+2\sqrt{2}m}r \ , \qquad \rho^2
\rightarrow \rho^2+ \frac{q m}{q+2\sqrt{2}m}r \ . \]
where $m$ denotes the black hole mass. Note that in both cases
$H$ (and $C(r,\theta)$ for the Zermelo case) changes accordingly by virtue of (\ref{H}).

\subsubsection{Ergo-region Randers/Zermelo structure}
\begin{figure}[h!]
\centering\includegraphics[height=3in,width=3in]{{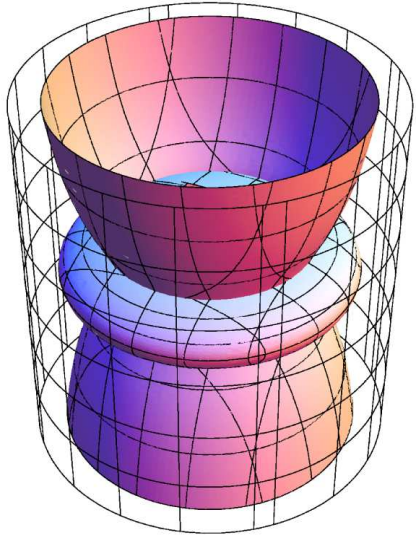}}
\caption{Illustration of the ergo-sphere and the break down surface of the Randers structure for the Kerr black hole with $a=0.98m$. The non-rotating Randers structure covers the region \textit{outside} the ergo-sphere. The co-rotating Randers structure covers the region \textit{inside} the cup-like surfaces. For $a/m$ obeying (\ref{intersect}) these surfaces intersect the ergo-sphere in the vicinity of the equatorial plane. They approach the light cylinder (meshed) asymptotically from within, for small co-latitudes. Note that this is \textit{not} an embedding diagram.}
\label{randerskerr1}
\end{figure}

In order to investigate the Randers structure up to the outer event horizon, we need to change to coordinates that are co-rotating with the outer event horizon of the Kerr black hole. Recall that the angular velocity of the outer horizon is 
\bequ
\Omega_+=\frac{a}{2mr_+} \ , \qquad r_{\pm}\equiv m\pm\sqrt{m^2-a^2} \ . \eequ
Thus one finds the suitable co-rotating angular coordinate (note that for $a>0$, freely falling observers in the geometry (\ref{kerr1}) are being dragged in the \textit{positive} $\phi$ direction)
\[
\phi'=\phi-\Omega_+ t\ .
\]
In these co-rotating Boyer-Lindquist coordinates, the line element of the Kerr solution can be written in the form (\ref{randersform}) with $V^2=H'/\rho^2$ and the Randers structure
\bequ
b'=\frac{N(r,\theta)\sin^2\theta}{H'}d\phi' \ , \qquad  a_{ij}'dx^idx^j=\frac{\rho^4}{H'}\left(\frac{dr^2}{\Delta}+d\theta^2+\frac{\Delta\sin^2\theta}{H'}d\phi'^2\right) \ ; \label{randerskerrnh}
\eequ
we have introduced
\begin{eqnarray*}
N(r,\theta)&\equiv& -2mra+\Omega_+ \frac{C(r,\theta)}{H} \ , \\
 H'&\equiv& H+4\Omega_+mra\sin^2\theta-\frac{\Omega_+^2\sin^2\theta C(r,\theta)}{H} \ .
\end{eqnarray*}
By inspection, one confirms that (\ref{randerskerrnh}) reduces to (\ref{randerskerr2}) when $\Omega_+=0$. This Randers structure breaks down when 
\[ |b'|^2=\Delta\Omega^2_+\sin^2\theta\left(1+\frac{2mr(r^2-r_+^2)}{\rho^2\Delta}\right)^2\ge 1 \ . \]
For non-extremal black holes $|b'|^2\rightarrow 0$ at the horizon, which confirms that this Randers structure is indeed valid near the horizon. Moving away from the horizon, $|b'|^2$ grows, except for $\theta=0$ and $\theta=\pi$, at which values $|b'|^2=0$ for any $r$. Clearly, the co-rotating Randers structure (\ref{randerskerrnh}) breaks down as $r\rightarrow\infty$. The boundary surface where this occurs depends both on $\theta$ and $r$, and it approximates $\Omega_+ r \sin \theta=1$ as $r\rightarrow\infty$, which is simply the light cylinder of a a rigidly rotating platform with the angular velocity of the outer Kerr horizon. Sufficiently close to extremality, the co-rotating Randers boundary surface enters the ergo-region of the black hole. By symmetry, this happens first at the equator as $a/m$ increases, and the condition for intersection is therefore
\bequ 
\frac{a}{m}\ge \sqrt{\sqrt{8}-2}\approx0.910 \ .\label{intersect}
\eequ
This is illustrated schematically in figure \ref{randerskerr1}.

Physically, this Randers structure can be interpreted as the rather involved magnetic field  $\mathbf{B'}=B'^r \partial_r+B'^\theta \partial_\theta$, where
\begin{eqnarray}
B'^r&=&\frac{2\cos\theta}{\rho^6}\left((H'+N\Omega_+\sin^2\theta)(N-\Omega_+\Delta a^2\sin^2\theta)+Na\sin^2\theta(a-2mr\Omega_+)\right)\ ,\label{kerrmagnetic2} \\
B'^\theta&=&\frac{2\sin\theta}{\rho^6}\left((H'+N\Omega_+\sin^2\theta)\left(ma-\Omega_+(2r(r^2+a^2)-(r-m)a^2\sin^2\theta)\right)\right.\nonumber \\
& & \mbox{}\left.+N(r-m+\Omega_+ ma\sin^2\theta)\right) \ . \nonumber
\end{eqnarray}
In this form, all $\Omega_+$ terms derive from the transformation to the co-rotating frame, so that the equations reduce to the Boyer-Lindquist case (\ref{kerrmagnetic}) if one sets $\Omega_+=0$. Note also that $B'^r=0$ on the equatorial plane, as expected from the fact that this is a totally geodesic sub-manifold.

As before, we can also give expressions for the Zermelo structure corresponding to this co-rotating Randers structure,
\begin{eqnarray*}
\mathbf{W'}&=& -\frac{N(r,\theta)H'}{\rho^4\Delta-N(r,\theta)^2\sin^2\theta}\frac{\partial}{\partial \phi'}\ , \\
h'_{ij}dx^idx^j&=&\lambda' \frac{\rho^4}{H'}\left(\frac{dr^2}{\Delta}+d\theta^2+\lambda'\frac{\Delta\sin^2\theta}{H'}d\phi'^2\right)\ ,
\end{eqnarray*}
using (\ref{zermelo1}) and (\ref{zermelo2}), where
\[
\lambda'=1-\frac{N(r,\theta)^2\sin^2\theta}{\rho^4\Delta}\ .
\]

\subsubsection{Near horizon for non extremal case: Asymptotic Hyperbolicity}
To analyse the near horizon Randers structure take
\[
r=r_+ + \epsilon \ , \ \ \mbox{and} \ \ \epsilon \rightarrow 0 \ .
\]
For the non-extremal case ($r_+>r_-$), as $\epsilon \rightarrow 0$, 
\[H'(\epsilon,\theta)= \frac{(r_+-r_-)}{(a^2+r_+^2)^2}\rho_+^4\epsilon +\mathcal{O}(\epsilon^2) \ , \qquad \rho_+^2\equiv r_+^2+a^2\cos^2\theta \ . \] 
It follows that the Randers structure becomes, to leading order in $\epsilon$
\bequ
b'=\frac{a\sin^2\theta(a^2+r_+^2)}{\rho_+^2}\left(1+\frac{2r_+^2}{\rho_+^
  2}\frac{r_++r_-}{r_+-r_-}\right)d\phi' \ , \eequ
\begin{equation}
a_{ij}'dx^i dx^j=\frac{(a^2+r_+^2)^2}{(r_+-r_-)^2}\frac{d\epsilon^2}{\epsilon^2}+\frac{(a^2+r_+^2)^2}{r_+-r_-}\frac{d\theta^2}{\epsilon}+\frac{(a^2+r_+^2)^4\sin^2\theta}{(r_+-r_-)\rho^4_+}\frac{d\phi'^2}{\epsilon} \ .
\label{randers3}
\end{equation}
The magnetic field is
\[
{\bf B}'=\epsilon^2\frac{2a\cos\theta (r_+^2-a^2)}{\rho_+^4(r_+^2+a^2)}\left(3-\frac{a^2}{r_+^2}\cos^2\theta\right)\frac{\partial}{\partial \epsilon} \ . \]
At leading order in $\epsilon$ there is only a radial component of the
magnetic field, which vanishes on the equatorial plane. Clearly,
if higher orders of $\epsilon$ are taken into account, then a
component in the $\theta$ direction will also emerge.

It turns out that the 3-geometry described by $a_{ij}'$ (\ref{randers3}) is asymptotically hyperbolic. To exhibit this property, one can introduce a radial tortoise coordinate $r^*$, as usual by defining $dr^{*2}=a_{\epsilon\epsilon}'d\epsilon^2$. Hence, as $\epsilon \rightarrow 0, \ r^* \rightarrow -\infty$, and integration yields
\[
\epsilon=k \exp\left(r^*\frac{r_+-r_-}{a^2+r_+^2}\right) \ ,
\]
where we choose the constant $k=r_+-r_-$. Notice that the exponent becomes $r^*/2m$ and $k=2m$ if $a=0$, as in the standard Regge-Wheeler coordinate for the Schwarzschild solution. The line element (\ref{randers3}) may now be written more concisely by redefining the tortoise coordinate in the following way,
\[
\eta=-r^*\frac{r_+-r_-}{2(a^2+r_+^2)}=-\frac{1}{2}\ln \frac{\epsilon}{r_+-r_-} \ ,
\]
such that the asymptotically hyperbolic structure becomes apparent,
\begin{equation}
a_{ij}'dx^i dx^j|_{\epsilon\rightarrow 0}=C^2(d\eta^2+\sinh^2\eta ds_2^2)|_{\chi\rightarrow \infty} \ ,
\label{randers4}
\end{equation}
where the conformal factor is
\[
C=2\frac{a^2+r_+^2}{r_+-r_-}\ ,
\]
and the line element of the hyperbolic boundary surface is
\bequ
ds_2^2=d\theta^2+\frac{(a^2+r_+^2)^2}{\rho^4(r_+)}\sin^2\theta d\phi'^2 \ , 
\label{hyperbboundary}
\eequ
which, for the Schwarzschild metric, reduces to the line element of the unit sphere.

It is instructive to compare the geometry of this boundary surface with the geometry of the event horizon at $r_+$. To this end, we introduce a distortion parameter $\beta\equiv a/\sqrt{r_+^2+a^2}$ as discussed by Smarr \cite{smarr}, where 
$0\leq \beta < 1/\sqrt{2}$ in the non-extremal case. Then, for small $\beta$, the surface is an oblately deformed sphere which can be 
globally embedded in $\mathbb{E}^3$. Moreover, one finds that the Gaussian curvature at the poles 
vanishes when $\beta=1/2$ or, if the black hole is un-charged, $a/m=\sqrt{3}/2$. 
For $\beta>1/2$, there are two polar caps of negative Gaussian curvature, and the event horizon cannot be 
globally embedded in $\mathbb{E}^3$ in this case. 

The surface described by the line element (\ref{hyperbboundary}) 
shows similar behaviour. To see this, consider its Gaussian curvature,
\begin{equation}
K(\theta)=1+\frac{2a^2}{\rho_+^4(\theta)}\left(r_+^2-(4r_+^2+3a^2)\cos^2\theta\right) \ .
\label{hyperbbgauss}
\end{equation}
Clearly, for $a\rightarrow 0$, one recovers the line element of the unit sphere with $K=1$. As $\beta$ increases, two polar 
caps of negative Gaussian curvature appear and the surface cannot be globally embedded in Euclidean space, as before. 
From (\ref{hyperbbgauss}), the transition occurs at $\beta=1/\sqrt{6}$, that is, $a/m=\sqrt{5}/3$. Furthermore, the topology of 
the boundary surface can be found from the Gauss-Bonnet theorem. It implies that the Euler characteristic is
\begin{eqnarray*}
\chi=\frac{1}{2\pi}\int K(\theta) e^\theta \wedge e^{\phi'} = 2 \ ,
\end{eqnarray*}
using (\ref{hyperbbgauss}) and an orthonormal dyad $\{e^\theta, e^{\phi'}\}$ on (\ref{hyperbboundary}). Hence, the 
boundary surface is a topological sphere, as in the case of the Kerr event horizon \cite{smarr}.
 
\subsubsection{Near horizon for extremal case: the NHEK geometry}
The Near Horizon geometry of an Extreme Kerr black hole, so called NHEK geometry \cite{Bardeen:1999px},  is particularly interesting. Recently \cite{Guica:2008mu} it has been argued that it is dual to a Conformal Field Theory and the suggested duality has been used to exactly compute the Bekenstein-Hawking entropy of an extreme Kerr black hole from the CFT. In global dimensionless coordinates $(\tau,r,\theta,\varphi)$ this Ricci flat geometry reads\footnote{The radial coordinate here $r$ relates to the global radial coordinate used in \cite{Bardeen:1999px,Guica:2008mu}, $r_{bh}$ by the simple transformation $r_{bh}=\sinh r$.}
\bequ ds^2=2j C^2(\theta)\left[-\cosh^2 r d\tau^2+dr^2+d\theta^2+\Lambda(\theta)^2(d\varphi+\sinh r d\tau)^2\right] \ , \label{nhek} \eequ
where
\[ C^2(\theta)\equiv \frac{1+\cos^2\theta}{2} \ , \qquad \Lambda(\theta)\equiv \frac{2\sin\theta}{1+\cos^2\theta} \ . \]
These global coordinates, with the ranges $-\infty<\tau,r<+\infty$, cover the whole spacetime.  For $\theta=0$ and $\theta=\pi$, the geometry is exactly $AdS_2$. There are two timelike conformal boundaries at $r=\pm \infty$. Thus, this geometry shares some properties of the $AdS_2\times S^2$ near horizon geometry of extreme charged black holes. The $\theta=constant$ sections have isometry group $SL(2,\mathbb{R})_L\times U(1)_R$, which is enhanced to $SL(2,\mathbb{R})_L\times SL(2,\mathbb{R})_R$ when $\Lambda(\theta)^2=1$. In the latter case the geometry is exactly $AdS_3$ and, inspection of (\ref{nhek}) reveals that it is written in an analytic continuation of Euler coordinates for $S^3$. 

The Randers structure for the NHEK geometry is
\bequ
\barr{c}
\displaystyle{b=-\frac{\Lambda(\theta)^2\sinh r}{\cosh^2r-\Lambda(\theta)^2\sinh^2r}d\varphi \ ,} \spa{0.3cm}\\ \displaystyle{ a_{ij}dx^idx^j=\frac{1}{\cosh^2r-\Lambda(\theta)^2\sinh^2r}\left(d\theta^2+dr^2+\frac{\Lambda(\theta)^2\cosh^2r}{\cosh^2r-\Lambda(\theta)^2\sinh^2r}d\varphi^2\right) \ .} \earr \eequ
One observes that the breakdown condition reduces to 
\[ |\coth r|=\Lambda(\theta) \ . \]
Thus the Randers structure is valid for \textit{all} $r$ if $2\sin\theta>1+\cos^2\theta$ that is inside a ``light-cone'' with polar angle $\theta=47^o$. Observe also that the induced metric on this ``light cone'' $\Lambda(\theta)^2=1$ is exactly that of $\mathbb{H}^2$ and $db$ becomes the volume form of  $\mathbb{H}^2$. The corresponding magnetic field is quite simple:
\[ {\bf B}=-\frac{2\cos\theta(2+\sin^2\theta)}{(1+\cos^2\theta)^2}\sinh 2r\frac{\partial}{\partial r}-\Lambda(\theta)(\cosh^2r-\Lambda(\theta)^2\sinh^2r-2)\frac{\partial}{\partial \theta} \ . \]

Inspection of (\ref{nhek}) immediately allows us to derive the Zermelo structure. Introducing a new radial coordinate $R\equiv \int dr/\cosh r$, with range $0<R<\pi$, the Zermelo structure becomes remarkably simple:
\bequ
{\bf W}=-\cot R\frac{\partial}{\partial \varphi} \ , \qquad h_{ij}dx^idx^j=dR^2+\sin^2R\left(d\theta^2+\Lambda(\theta)^2d\varphi^2\right) \ . \label{zermelonhek}\eequ
The 2-geometry $d\theta^2+\Lambda(\theta)^2d\varphi^2$ is exactly the $a=r_+$ limit of (\ref{hyperbboundary}). In particular it is topologically a 2-sphere. Thus, the 3-geometry in (\ref{zermelonhek}) is, topologically, a 3-sphere. Hence, this Zermelo structure describes an azimuthal wind, varying with R, in a deformed $S^3$.

\section{Concluding Remarks}
\label{final}
In this paper we have explored a triangle of relations  between stationary spacetimes $\{\mathcal{M}, g\}$ and two structures on the space of orbits $\mathcal{B}$ of the time translations Killing vector field ${\bf K}$: a \textit{Randers-Finsler structure} $\{\mathcal{B}, a_{ij},b_i\}$ and a \textit{Zermelo structure} $\{\mathcal{B}, h_{ij},W^i\}$. This triangle can be seen as a triality between physical problems: the null geodesic flow in $\{\mathcal{M}, g\}$, the magnetic flow due to $db$ on $\{\mathcal{B},a_{ij}\}$ and the Zermelo navigation problem on $\{\mathcal{B},h_{ij}\}$ due to the wind ${\bf W}$. But each of these physical problems motivates alternative viewpoints.

From the spacetime perspective, the study of null geodesics motivates the following question:  what generalises the usual optical geometry for the case of stationary, rather than static, spacetimes? If we wish to keep a strictly geometric viewpoint the answer is, as we have described, a Randers-Finsler, rather than Riemannian, geometry. A connection along these lines had already been hinted in \cite{bj}. This viewpoint provided a deep insight into the construction of Randers-Finsler geometries of constant flag curvature: they all correspond to conformally flat spacetimes. It would be interesting to prove the converse statement. Also of interest might be to write down tensorial relations, in the Randers and Zermelo pictures, for circumstances in which the magnetic field/wind or the space curvature might be gauged away, by moving to an equaivalent problem. 

Interestingly, as we have shown in the Kerr example, the Riemannian part of the Randers-Finsler geometry still approaches hyperbolic space in the vicinity of a rotating, non-degenerate Killing horizon, as for the case of non-rotating ones. One should be able to show that this is a generic feature for rotating, non-degenerate Killing horizons and generalise some of the applications  \cite{GibbonsWarnick} given in the non-rotating case. 

This connection also allows a reinterpretation of the Finsler condition from the spacetime viewpoint. As we have argued and shown in a number of examples, the break down of the Randers condition is associated to either the change in character of ${\bf K}$, as on an ergosphere, or to the non-existence of a global time function associated to  ${\bf K}$, as in situations where ctc's arise.

Taking a more physical interpretation, rather than a strictly geometric one, a Randers-Finsler flow is simply a magnetic flow in a Riemannian geometry. One may then recast properties of the former in the language of the latter. For instance, in three dimensions, all constant flag curvature Randers-Finsler geometries are Killing magnetic fields. Another connection of great beauty is that provided by squashed $SL(2,\mathbb{R})$ example: the existence of a break down surface for the Randers structure is associated to a change in the ergodicity properties of the magnetic flow.

We have described how the connection between the spacetime and the Zermelo viewpoints arises by casting the former in a Painlev\'e-Gullstrand form. This is also the natural form for the spacetime metrics of analogue models, which conceptually also fit well with the Zermelo picture. 

An interesting open question is if this network of relations might be generalised if one allows a \textit{time dependent} wind in the Zermelo picture and how such generalisation works.  

\section*{Acknowledgements}
C.H. would like to thank the hospitality of D.A.M.T.P., University of
Cambridge, where part of this work was done and
financial support from the Centre of Theoretical Cosmology. C.M.W. thanks Queens' College, Cambridge for a Research Fellowship and is also grateful to
Centro de Fisica do Porto for hospitality during the preparation of this paper. M.C.W. gratefully acknowledges funding from S.T.F.C., United Kingdom.

\end{document}